\documentclass[aps,amsfonts,twocolumn,nofootinbib,floatfix]{revtex4}
\usepackage{amsmath}
\usepackage{epsfig}
\usepackage{latexsym}
\usepackage{color}
\usepackage{pstricks}

\begin{document}

\title{Brane Universes with Gauss-Bonnet-Induced-Gravity}

\author{Richard A. Brown$^{1}$}

\affiliation{\vspace*{0.2cm} $^1$Institute of Cosmology \&
Gravitation, University of Portsmouth, Portsmouth~PO1~2EG, UK}

\date{3 May 2006}

\begin{abstract}
The DGP brane world model allows us to get the observed late time
acceleration via modified gravity, without the need for a ``dark
energy'' field. This can then be generalised by the inclusion of
high energy terms, in the form of a Gauss-Bonnet bulk. This is the
basis of the Gauss-Bonnet-Induced-Gravity (GBIG) model explored here
with both early and late time modifications to the cosmological
evolution. Recently the simplest GBIG models (Minkowski bulk and no
brane tension) have been analysed. Two of the three possible
branches in these models start with a finite density ``Big-Bang''
and with late time acceleration. Here we present a comprehensive
analysis of more general models where we include a bulk cosmological
constant and brane tension. We show that by including these factors
it is possible to have late time phantom behaviour.

\end{abstract}

\maketitle

\section{Introduction}

There are some intriguing problems with the standard $\Lambda$CDM
cosmology that may hint at new physics in order to answer them.
One of the problems is the requirement of adding an additional
``dark energy'' field in order to explain the acceleration that
the universe is now experiencing. An alternative way of explaining
this phenomenon is to modify gravity at large scales. This should
not be done in an ad hoc way but requires a consistent and
covariant physical grounding. The brane model put forward by
Dvali, Gabadadze and Porrati (DGP)~\cite{Dvali:2000hr}, and
generalised to cosmological branes by
Deffayet~\cite{Deffayet:2000uy}, has such a grounding.

Brane models are inspired by the discovery of D-branes within
string theories. The key property is that matter is confined to
the brane. The DGP model has a large scale/low energy effect of
causing the expansion rate of the universe to accelerate. This is
achieved via the addition of an Induced-Gravity (IG) term to the
gravitational action. The IG term is produced by the quantum
interaction between the matter confined on the brane and the bulk
gravity. It has also been shown that the IG term can be obtained,
in some string theory models, from the presence of Gauss-Bonnet
gravity in the bulk~\cite{Mavromatos:2005yh}.

One of the problems with the DGP model is the fact that it is in
some sense ``unbalanced''. This is because it does not modify the
short-distance/high-energy regime which we would expect from a
string and quantum motivated theory. Brane models with Gauss-Bonnet
(GB)~\cite{gb} bulk gravity on the other hand modify the high-energy
regime like the original Randall and Sundrum (RS)
models~\cite{Randall:1999vf} did. The RS models have early times
described by 5D gravity because the bulk is warped. This warping
localises gravity to the brane in the low energy regime while in the
high energy era gravity leaks off the brane and ``sees'' the bulk
and thus acts 5D. The DGP model has 5D behaviour once a certain
length scale (the cross-over scale) has been reached. At this scale
gravity starts to leak off the brane and into the bulk thus causing
gravity to become strongly 5D. If we try warping the bulk in the DGP
model with the aim of achieving both an early-time and a late-time
modification, we ultimately fail. It has been shown~\cite{co} that
the DGP model in a warped bulk either has 4D gravity behaviour at
all scales or has 4D gravity at short and long scales with 5D
behaviour at intermediate scales.

We include GB gravity in the bulk with IG on the brane
(GBIG)~\cite{Kofinas:2003rz,Cai:2005ie} with the goal of obtaining a
model that is ``balanced'', i.e. a model that gives us both UV and
IR modifications.

In Ref.~\cite{Brown:2005ug} we looked at the simplest generalization
of the DGP model. We saw that this model provides us with both early
and late time modification of GR. We found that by including both
the GB and IG terms the ``big-bang'' singularity can have a finite
density. In this work we systematically investigate the other
solutions allowed in the model, i.e. we no longer restrict ourselves
to a Minkowski bulk with zero brane tension.

\section{Field Equations}

The general gravitational action contains the Gauss-Bonnet (GB)
term in the bulk and the Induced Gravity (IG) term on the ($Z_2$
symmetric) brane:
\begin{eqnarray}\label{AcGBIG}
S_{\rm grav}&=&\frac{1}{2\kappa_5^2}\int d^5 x\sqrt{-
g^{(5)}}\left\{ R^{(5)} -2\Lambda_5
\right.\nonumber\\&+&\left.\alpha\left[ R^{(5)2}-4
R^{(5)}_{ab}R^{(5)ab}+ R^{(5)}_{abcd}R^{(5)abcd}\right]\right\}
\nonumber\\&+&\frac{r}{2\kappa^2_5} \int_{\rm brane}d^4x\sqrt{-
g^{(4)}}\,\left[R^{(4)}-2\frac{\kappa^2_5}{r}\lambda\right]\,,
\end{eqnarray}
where $r\geq0$ is the Induced Gravity `cross-over' scale
($r=\kappa^2_5/\kappa^2_4$ consistent with
Ref.~\cite{Kofinas:2003rz,Brown:2005ug}), $\alpha$ is the
Gauss-Bonnet coupling constant and $\lambda$ is the brane tension. A
negative value of $\alpha$ would require the string coupling
constant to be imaginary. We could still use the GB gravity in a
classical sense but we would lose our string motivation.
\begin{figure}
\includegraphics[height=3in,width=2.75in,angle=270]{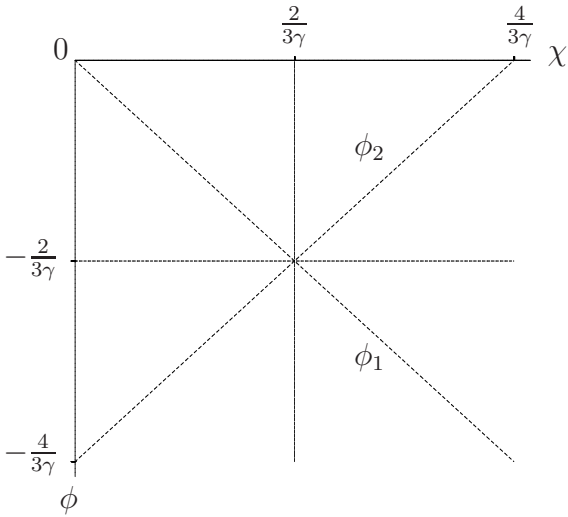}
\rput(-0.5,-0.7){\large$\chi$}
\rput(-4,-0.3){\large$\frac{2}{3\gamma}$}\rput(-1,-0.3){\large$\frac{4}{3\gamma}$}
\rput(-7,-6.6){\large $\phi$} \rput(-7.1,-0.6){\large$0$}
\rput(-7.5,-6){\large $-\frac{4}{3\gamma}$}
\rput(-7.5,-3.4){\large $-\frac{2}{3\gamma}$}
\rput(-3,-1.9){\large $\phi_2$} \rput(-3,-4.7){\large$\phi_1$}
\caption{The effective bulk cosmological constant $\phi$ as a
function of $\chi$.} \label{bulkC}
\end{figure}
We have the standard conservation equation:
\begin{equation}\label{ec}
\dot \rho+3H(1+w)\rho=0\,,~w=p/\rho\,,
\end{equation}
so the matter that is on the brane does not exchange energy with
the bulk and acts as a perfect fluid as in normal Friedmann
solutions. The general Friedmann equation was obtained in
Ref.~\cite{Kofinas:2003rz}. Here we assume that the bulk black
hole mass is zero and the brane is spatially flat, but we allow
non-zero brane tension and bulk curvature. The general form of the
Friedmann equation is then:

\begin{equation}\label{Fried}
4\left[1+\frac{8}{3}\alpha\left(H^2+\frac{\Phi}{2}\right)\right]^2\left(H^2-\Phi\right)
=\left[rH^2-\frac{\kappa^2_5}{3}(\rho+\lambda)\right]^2,
\end{equation}
where $\Phi$ is a solution to:

\begin{equation}\label{Phi}
\Phi+2\alpha\Phi^2=\frac{\Lambda_5}{6}.
\end{equation}
The bulk cosmological constant is given by,
Ref.~\cite{Dufaux:2004qs}:

\begin{equation}\label{Lamb5}
\Lambda_5=-\frac{6}{\ell^2}+\frac{12\alpha}{\ell^4}.
\end{equation}
Equations~(\ref{Phi}) and (\ref{Lamb5}) give us two solutions for
$\Phi$:

\begin{equation}\label{Phi2}
\Phi_1=-\frac{1}{\ell^2},~~\Phi_2=\frac{1}{\ell^2}-\frac{1}{2\alpha}.
\end{equation}

We work with the Friedmann equation in dimensionless form by
defining the following variables (with $r>0$):
\begin{eqnarray}\label{DV}
\gamma&=&\frac{8\alpha}{3r^2},~ h=Hr,~\mu
=\frac{r\kappa^2_5}{3}\rho,~ \sigma
=\frac{r\kappa^2_5}{3}\lambda,\nonumber\\\chi&=&\frac{r^2}{\ell^2},
~\phi=\Phi r^2,~\tau=\frac{t} {r}.
\end{eqnarray}
Using these variables we can write the two solutions for $\Phi$ as:

\begin{equation}\label{Phi3}
\phi_1=-\chi,~~\phi_2=\chi-\frac{4}{3\gamma}.
\end{equation}

The bulk cosmological constant gives us an upper bound on the GB
coupling constant $\alpha$. Equation~(\ref{Lamb5}) gives us:

\begin{equation}\label{lcon}
\frac{1}{\ell^2}=\frac{1}{4\alpha}\left[1\pm\sqrt{1+\frac{4}{3}\alpha\Lambda_5}\right].
\end{equation}
For an RS ($\alpha\rightarrow0$) limit we take the minus branch.
$\Lambda_5>0$ would imply $\ell^2<0$ and thus we must have
$\Lambda_5\leq0$ and:

\begin{equation}\label{acon1}
\alpha\leq\frac{\ell^2}{4}.
\end{equation}
In dimensionless form this is given by:

\begin{equation}\label{alCon1}
\gamma\leq \frac{2}{3\chi}.
\end{equation}
Maintaining an RS limit would also rule out the $\phi_2$ branch. We
would be restricted to the solutions lying along the line in the top
left quadrant in Fig.~\ref{bulkC}. We are interested in the whole
range of the model so we include the plus branch in
Eq.~(\ref{lcon}). We assume $\Lambda_5\leq0$ therefore our
constraint on $\alpha$ is given by:

\begin{equation}\label{acon2}
\alpha\leq\frac{\ell^2}{2},~\Rightarrow~\gamma\leq \frac{4}{3\chi}.
\end{equation}
If we take $\chi=0$ then we have no bound on $\gamma$ (apart from
being positive and real). This is the case considered in
Ref.~\cite{Brown:2005ug}.

In Fig.~(\ref{bulkC}) we have the two $\phi$ solutions plotted as
functions of $\chi$. The two solutions with $\phi=0$ both live in a
Minkowski bulk, all the rest live in an AdS bulk. Note that one of
these AdS solutions ($\chi=0,\phi=-4/3\gamma$) has $\Lambda_5=0$ but
$\Phi=-1/2\alpha$ acting as an effective cosmological constant
\cite{Brown:2005ug}. We see that for any allowed value of $\phi$ we
can be on either of the two branches. This means that we need not
consider the $\phi_1$ and $\phi_2$ solutions to the Friedmann
equation separately. We therefore consider Eq.~(\ref{acon2}) in
terms of $\phi$. We define the maximum value of $\gamma$, for a
particular value of $\phi$, that's allowed by the constraint
equation:

\begin{equation}\label{gammaM}
\gamma_{\rm M}=-\frac{4}{3\phi}.
\end{equation}

The dimensionless Friedmann equation is:

\begin{equation}\label{dFried}
4\left[1+\gamma\left(h^2+\frac{\phi}{2}\right)\right]^2\left(h^2-\phi\right)
=\left[h^2-(\mu+\sigma)\right]^2.
\end{equation}
with the conservation equation now given by:
\begin{equation}\label{ecd}
\mu'+3h(1+w)\mu=0,
\end{equation}
where $'=d/d\tau$ and $h=a'/a$. The Raychaudhuri and acceleration
equations are given by:
\begin{widetext}
\begin{equation}\label{Ray}
h'=\frac{3\mu(1+w)[h^2-(\mu+\sigma)]}{4(\gamma h^2+1)(3\gamma
h^2+1)-2[h^2-(\mu+\sigma)]-\phi\gamma(4+3\phi\gamma)},
\end{equation}
and:

\begin{equation}\label{accel}
\frac{a''}{a}=\frac{4h^2(\gamma h^2+1)(3\gamma
h^2+1)-[h^2-(\mu+\sigma)][2h^2-3\mu(1+w)]-h^2\phi\gamma(4+3\phi\gamma)}{4(\gamma
h^2+1)(3\gamma
h^2+1)-2[h^2-(\mu+\sigma)]-\phi\gamma(4+3\phi\gamma)}.
\end{equation}
\end{widetext}

\section{Friedmann Equation Solutions}

\begin{figure}
\includegraphics[height=3in,width=2.75in,angle=270]{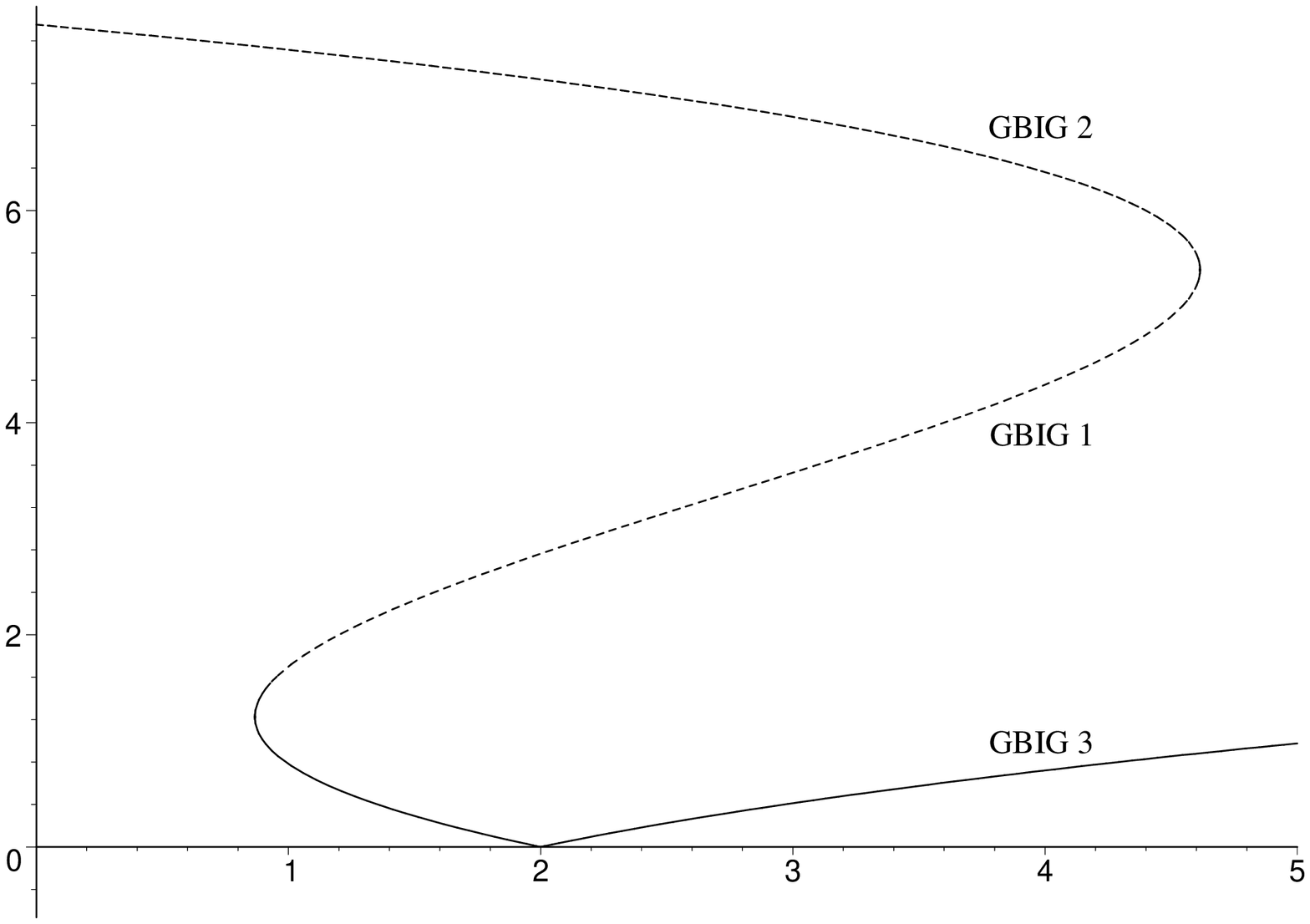}
\rput(-0.7,-2.3){\large$\mu_{\rm i},h_{\rm i}$}
\rput(-4.5,-5.6){\large$\mu_l$} \rput(-6.4,-5){\large $\mu_{\rm
e},h_{\rm e}$} \rput(-7.4,-0.8){\large$h_{\infty}$}
\rput{350}(-3.5,-1.25){\large $\leftarrow$}
\rput{20}(-3.5,-3.8){\large $\leftarrow$}
\rput{11}(-1.2,-5.2){\large $\leftarrow$}
\rput{335}(-5,-5.65){\large$\leftarrow$}\rput(-4,-6.2){\large
$\mu$} \rput(-7.4,-3.2){\large $h$}
 \caption{Solutions of the Friedmann equation
($h$ vs $\mu$) with negative brane tension ($\sigma=-2$) in a
Minkowski bulk ($\phi=0$) with $\gamma=1/20$. The curves are
independent of the equation of state $w$. The arrows indicate the
direction of proper time on the brane.} \label{Friedpl1}
\end{figure}
\begin{figure}
\includegraphics[height=3in,width=2.75in,angle=270]{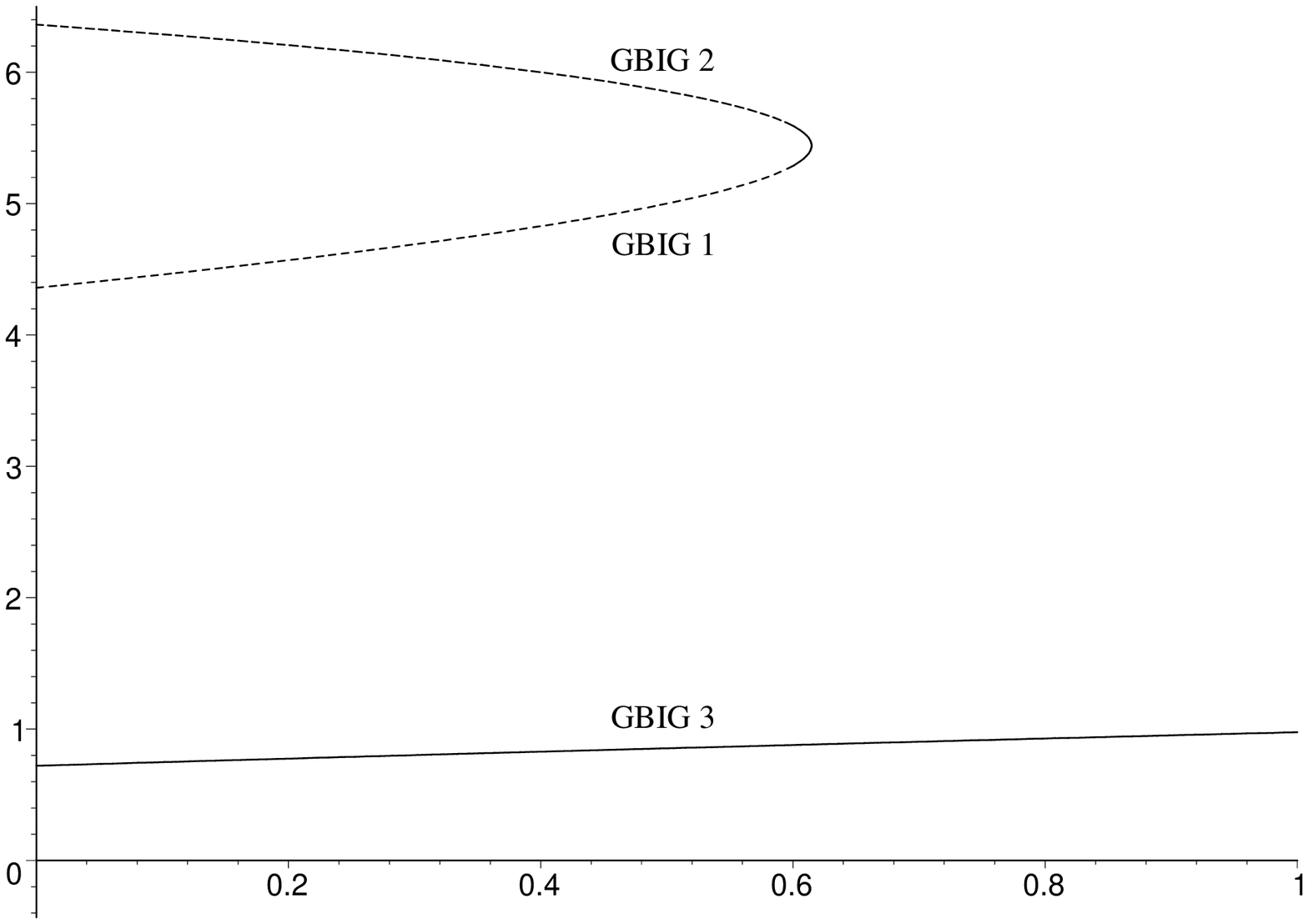}
\rput(-2.7,-1.6){\large$\mu_{\rm i},h_{\rm i}$}
\rput(-7.4,-0.8){\large $h_{\infty}$} \rput(-7.4,-5.4){\large
$h_{\infty}$}\rput(-7.4,-2.5){\large $h_{\infty}$}
\rput{350}(-5,-0.95){\large $\leftarrow$}
\rput{12}(-5,-2.25){\large $\leftarrow$}
\rput{5}(-1.2,-5.1){\large $\leftarrow$}\rput(-4,-6.2){\large
$\mu$} \rput(-7.4,-3.4){\large $h$}
 \caption{Solutions of
the Friedmann equation with positive brane tension ($\sigma=2$) in a
Minkowski bulk ($\phi=0$) with $\gamma=1/20$.} \label{Friedpl2}
\end{figure}

The model we considered in Ref.~\cite{Brown:2005ug} is the
Minkowski bulk limit ($\chi=0$) of the $\phi_1$ case, which we now
see to be equivalent to $\chi=\frac{4}{3\gamma}$ in the $\phi_2$
case. The results we present below are in terms of $\phi$ and thus
include both $\phi_1$ and $\phi_2$. We will consider the effect of
including brane tension in a Minkowski bulk before we look at the
AdS bulk cases. The Minkowski bulk case is given by $\phi=0$,
since Eqs.~(\ref{Lamb5}) and (\ref{Phi2}) imply $\Lambda_5=0$.
\subsection{Minkowski bulk ($\phi=0$) with brane tension}

In Figs.~\ref{Friedpl1} and~\ref{Friedpl2} we can see the solutions
to the Friedmann equation in a Minkowski bulk. For both negative and
positive brane tensions there are three solutions (this is not
always true as we will show later) denoted GBIG1-3. There are four
points of interest in Figs.~\ref{Friedpl1} and~\ref{Friedpl2}, these
are:
\begin{itemize}
\item ($\mu_{\rm i},h_{\rm i}$) and ($\mu_{\rm e},h_{\rm i}$): The initial
density for GBIG1-2 ($\mu_{\rm i}$) and the final density for GBIG1
and 3 ($\mu_{\rm e}$), found by considering $d\mu/d(h^2)=0$, are
given by:

\begin{equation}\label{mui}
\mu_{\rm
i,e}=\frac{1-18\gamma\pm(1-12\gamma)^{3/2}}{54\gamma^2}-\sigma,
\end{equation}
where the plus sign is for $\mu_{\rm i}$ and the negative sign is
for $\mu_{\rm e}$. The Hubble rates for these two densities are
given by:

\begin{equation}\label{Hie}
h_{\rm
i,e}=\frac{\sqrt{2}}{6\gamma}\sqrt{1-6\gamma\pm\sqrt{1-12\gamma}},
\end{equation}
where the sign convention is the same as above. The points
($\mu_{\rm i,e},h_{\rm i,e}$) have $h'=|\infty|$. Therefore
cosmologies that evolve to $\mu_{\rm e},~h_{\rm e}$ end in a
``quiescent'' (finite density) future singularity. This singularity
is of type 2 in the notation of Ref.~\cite{Shtanov:2002ek}.

\item ($\mu_{ l},0$): This is the density at which GBIG3 ``loiters''. The
point was found by considering $h=0$ and is given by :

\begin{equation}\label{muL}
\mu_{l}=-\sigma,
\end{equation}
In the case considered in Ref.~\cite{Brown:2005ug} we had $\sigma=0$
so $\mu_l=0$. We can show that for $\sigma<0$ GBIG3 will not
collapse but will loiter at a density of $\mu=-\sigma$ before
evolving towards ($\mu_{\rm e},h_{\rm e}$), by considering $h'_{
l}$. At $\mu=-\sigma$, $h=0$ we get $h'_{ l}=0$ from
Eq.~(\ref{Ray}). In a standard expanding or collapsing cosmology
$h'<0$ at all times. The evolution for a radiation dominated
universe can be seen in Fig.~\ref{Phi1GBIG3Mink1}. A dust dominated
universe spends longer at $\mu_l$.
\begin{figure}
\includegraphics[height=3in,width=2.75in,angle=270]{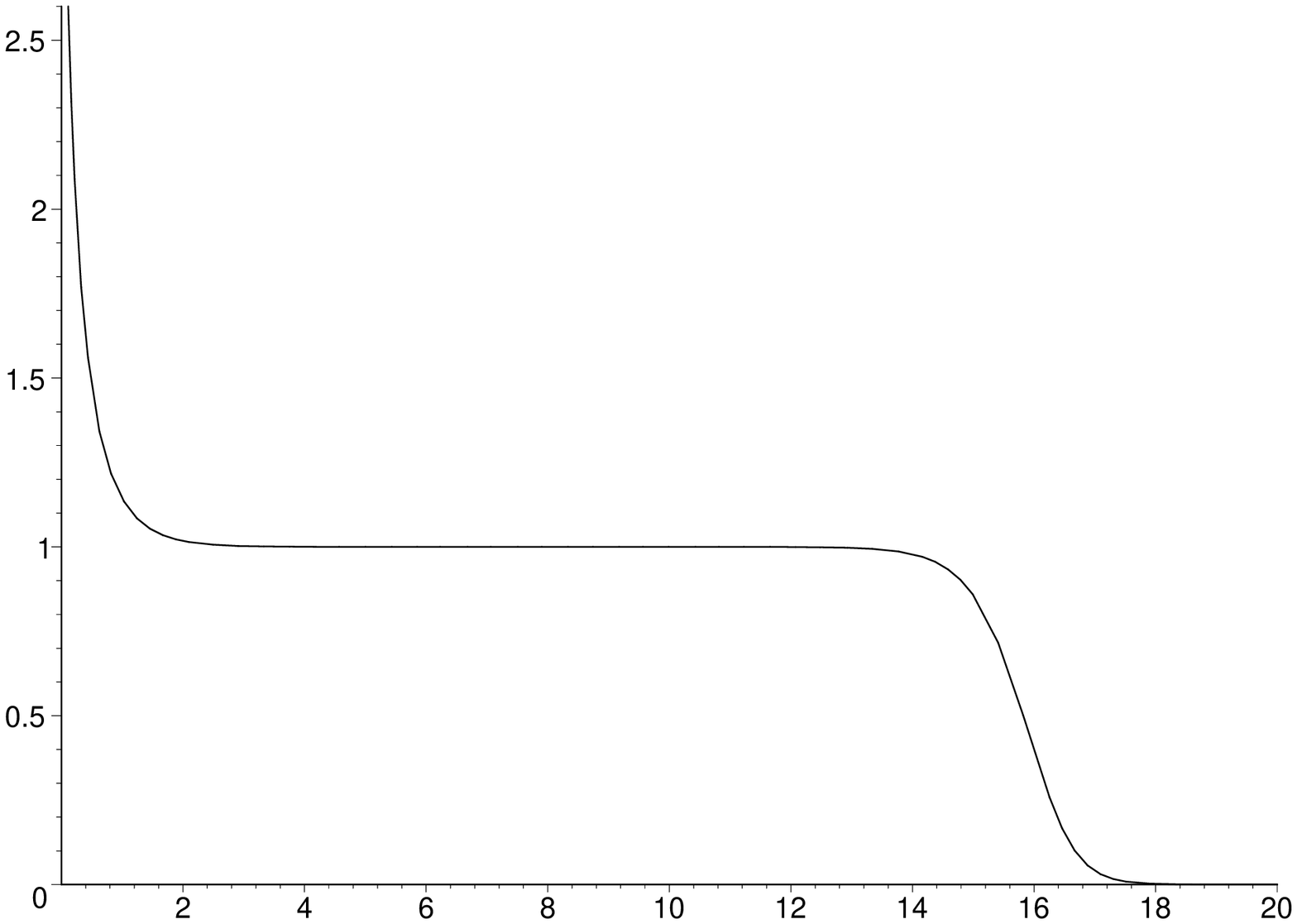}
\includegraphics[height=3in,width=2.75in,angle=270]{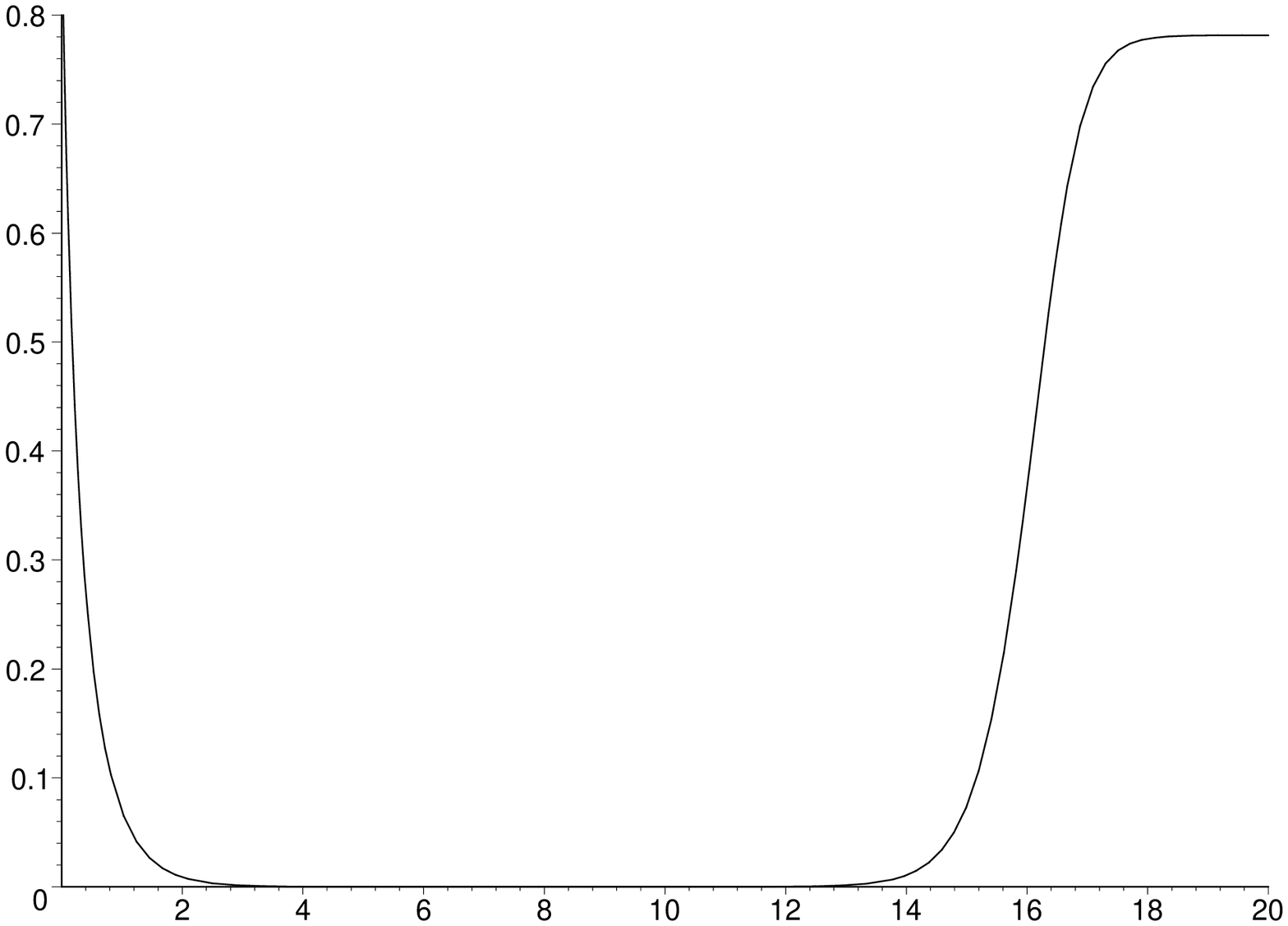}
\rput(-7.2,3.6){\large $\mu$} \rput(-4,0.3){\large
$\tau$}\rput(-7.5,-3.5){\large $h$}\rput(-4,-6.7){\large $\tau$}
 \caption{Plots of $\mu$ vs $\tau$ and
$h$ vs $\tau$ for GBIG3 in a Minkowski bulk ($\phi=0$) with
negative brane tension ($\sigma=-1$). We see that GBIG3 loiters
around $\mu=\mu_l=-\sigma$, $h=0$ before evolving towards a vacuum
de Sitter solution ($\mu_{\rm e}<0$). (Here $\gamma=1/20$ and
$w=1/3$.)} \label{Phi1GBIG3Mink1}
\end{figure}
In Ref.~\cite{Sahni:2004fb} they consider a loitering braneworld
model. An important difference between the two models is that in
Ref.~\cite{Sahni:2004fb} they require negative dark radiation,
i.e. a naked singularity in the bulk or a de Sitter bulk. Also in
our model the Hubble rate at the loitering phase is exactly zero.
The time spent at this point is only dependent on $w$ and the
density of the loitering phase is only dependent on the brane
tension.

\item ($0,h_{\infty}$):  This is the asymptotic value of the Hubble rate
as $\mu\rightarrow 0$. For the value of the parameters used in
Fig.~\ref{Friedpl1}, GBIG2 is the only case with this limit.  In
general any of the models can end in a similar state
(Fig.~\ref{Friedpl2}). The different values of $h_{\infty}$ are
given by the solutions to the cubic:
\begin{equation}\label{Phi1hin}
h_{\infty}^6+\frac{(8\gamma-1)}{4\gamma^2}h_{\infty}^4+\frac{2+\sigma}{2\gamma^2}h_{\infty}^2
-\frac{\sigma^2}{4\gamma^2}=0.
\end{equation}
In the case we considered in Ref.~\cite{Brown:2005ug} ($\sigma=0$)
the last term in the above cubic is zero. Therefore we have
$h_{\infty}=0$ for GBIG3 while the solutions for GBIG1-2 are given
by:

\begin{equation}\label{Phi100}
h_\infty=\frac{1} {2\gamma\sqrt{2}}\sqrt{1-8\gamma\mp
\sqrt{1-16\gamma}},
\end{equation}
where the minus sign corresponds to GBIG1 and the plus sign to
GBIG2. This is the only case where we can write simple analytic
solutions as in all other cases we have to solve the cubic.
\end{itemize}

\begin{figure}
\includegraphics[height=3in,width=2.75in,angle=270]{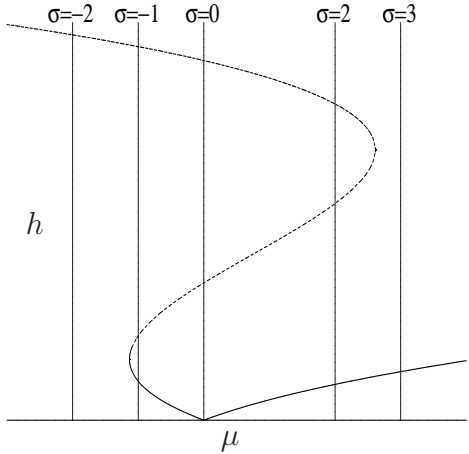}
\rput(-4,-6.4){\large $\mu$} \rput(-6.6,-3.5){\large $h$}
\caption{Solutions of the Friedmann equation ($h$ vs $\mu$) in a
Minkowski bulk ($\phi=0$) with $\gamma=1/20$. The vertical lines
represent the $\mu=0$ axis for the labeled brane tensions.}
\label{Phi1nszc}
\end{figure}
The two Hubble rates, $h_{\rm i}$ and $h_{\rm e}$, are independent
of the brane tension which simply shifts the $\mu=0$ axis.
Therefore $\mu_{\rm e}$ only corresponds to a physical (positive)
energy density when $\sigma<\sigma_{\rm e}<0$, see
Eq.~(\ref{sil}). The effect of the brane tension on the $\mu=0$
axis is illustrated in Fig.~\ref{Phi1nszc}.

\begin{figure}
\includegraphics[height=3in,width=2.75in,angle=270]{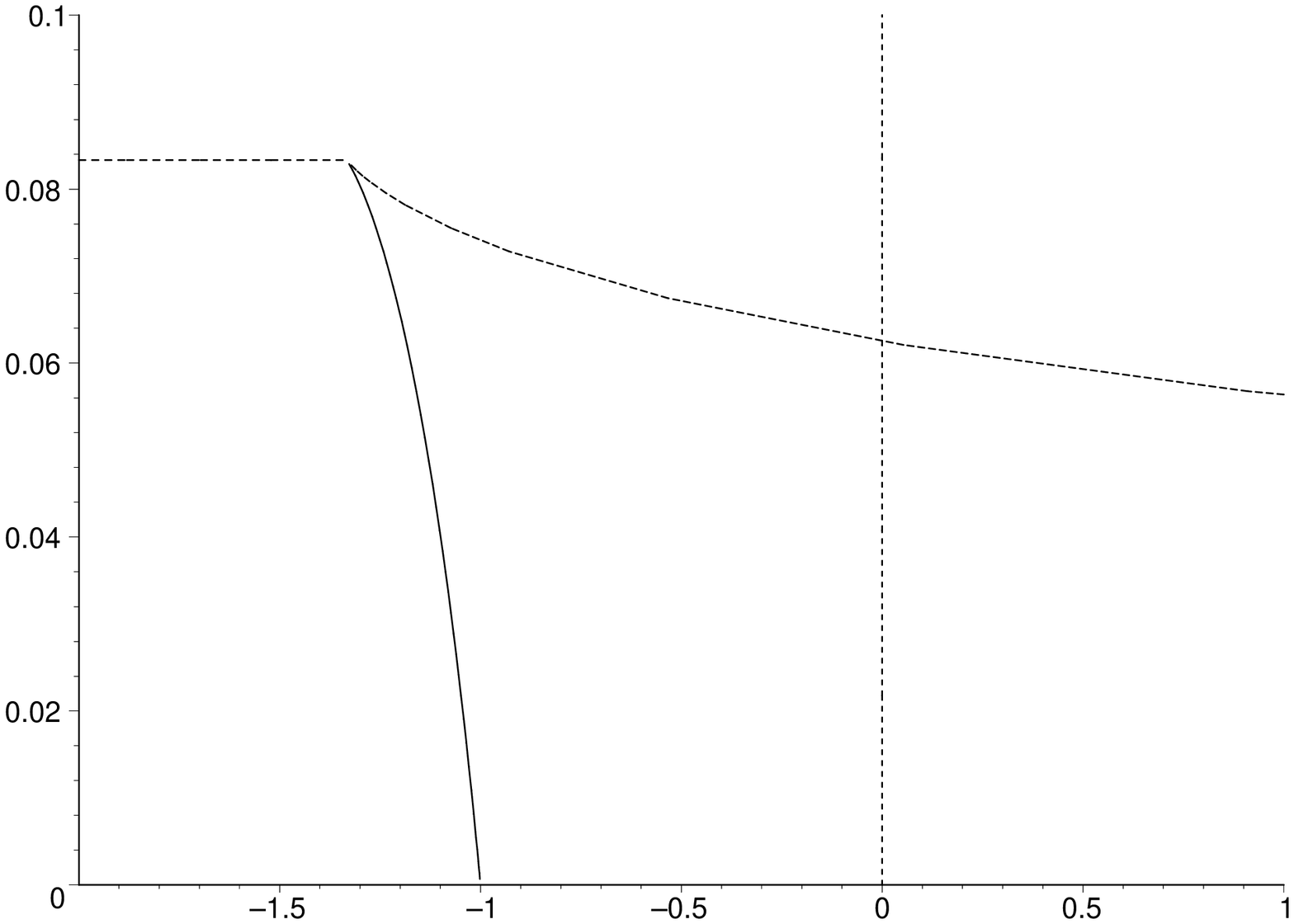}
\rput(-4.5,-1.25){\large I}\rput(-1.8,-1.25){\large
II} \rput(-6,-4.5){\large III} \rput(-4,-4.5){\large IV}
\rput(-1.8,-4.5){\large V}
 \rput(-5.7,-3.4){\large $\sigma_{\rm
e}(\gamma)$} \rput(-1,-3.4){\large $\gamma_{\rm
i}(\sigma)$}\rput(-3.1,-6){\large $\sigma_l$}
\rput(-6.2,-1.4){\large $\gamma_{\rm m}$}\rput(-7.2,-3.5){\large
$\gamma$}\rput(-3.9,-6.7){\large $\sigma$} \caption{The
$(\sigma,\gamma)$ plane for solutions in a Minkowski ($\phi=0$)
bulk. The short dotted horizontal line is $\gamma_{\rm m}=1/12$.
GBIG1-2 exist with positive energy density in regions III, IV and
V.} \label{gs0}
\end{figure}

We showed in Ref.~\cite{Brown:2005ug} that for $\phi=0=\sigma$ we
require $\gamma\leq1/16$ for GBIG1-2 solutions to exist within the
positive energy density region (if $\gamma=1/16$ GBIG1-2 reduce to
the same vacuum de Sitter universe). If we have some non-zero brane
tension this constraint is modified. The maximum value of $\gamma$,
for GBIG1-2 to exist with positive energy density, as a function of
$\sigma$ can be seen in Fig.~\ref{gs0}. We have defined new
quantities, $\sigma_{\rm e,i}$ and $\sigma_l$, for which $\mu_{\rm
e,i,l}=0$. We can see from Eq.~(\ref{muL}) that $\sigma_l=0$.
$\sigma_{\rm e,i}$ are given in terms of $\gamma$ by:

\begin{equation}\label{sil}
\sigma_{\rm i,e}=\frac{1-18\gamma\pm(1-12\gamma)^{3/2}}{54\gamma^2}.
\end{equation}

The maximum value of $\gamma$ ($\gamma_{\rm m}$) for GBIG1-2 to
exist (i.e. for Eqs.~(\ref{mui}) and (\ref{Hie}) to have real
solutions) is $\gamma_{\rm m}=1/12$ . Actually for $\gamma=1/12$,
GBIG2 exists but GBIG1 is lost. This is since $h_{\rm i}=h_{\rm e}$
at $\gamma=1/12$. The point $h_{\rm i}=h_{\rm e}$ is now a point of
inflection and Eq.~(\ref{Hie}) is still valid and therefore
$|h'|=\infty$. For values of $\gamma>1/12$, $h_{\rm i,e}$ and
$\mu_{\rm i,e}$ become complex, and it is no longer a point of
inflection. Therefore $|h'|\neq\infty$ and GBIG3 can continue its
evolution through this point to $h_{\infty}$.
\begin{figure}
\includegraphics[height=3in,width=2.75in,angle=270]{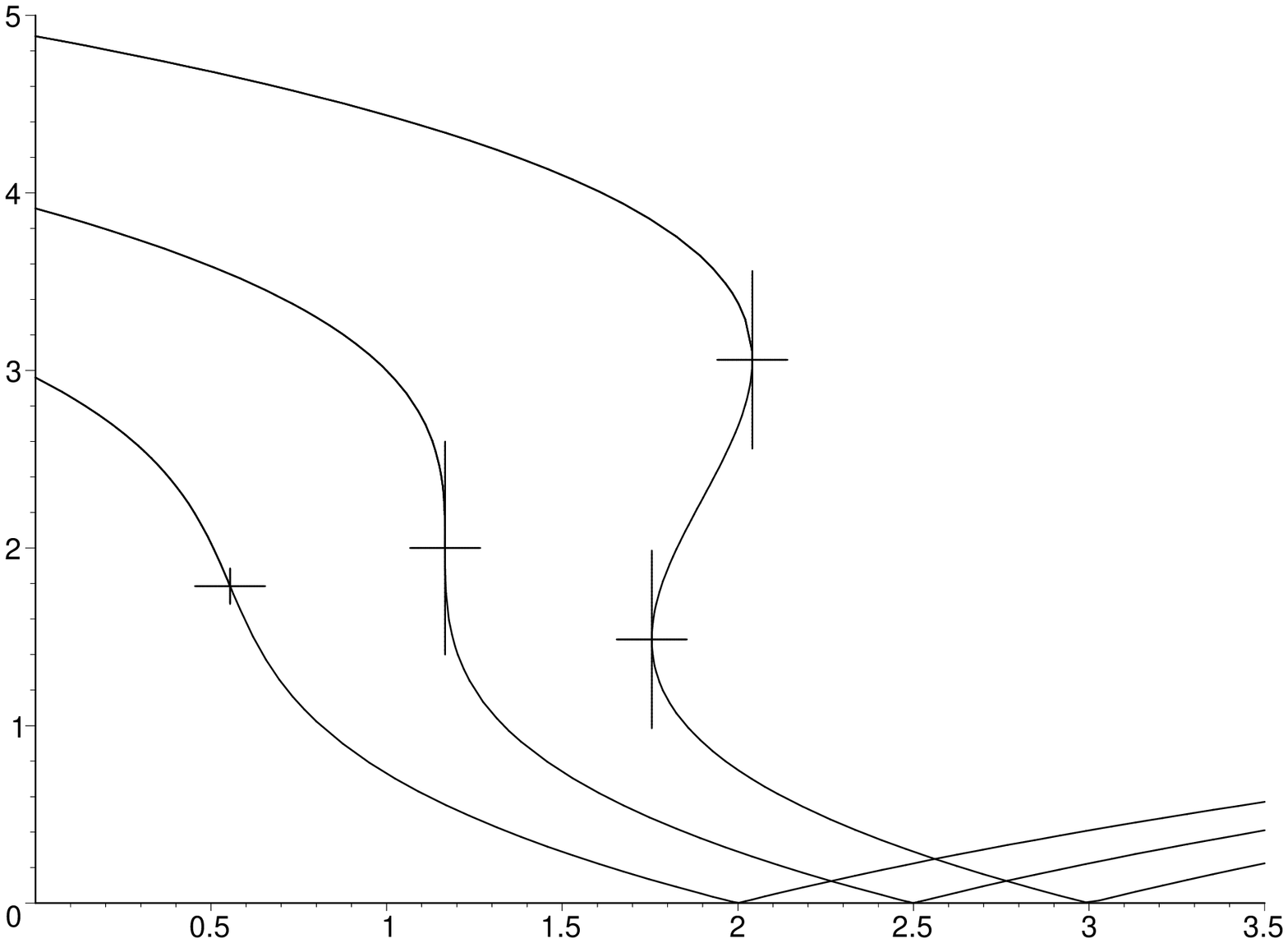}
\includegraphics[height=3in,width=2.75in,angle=270]{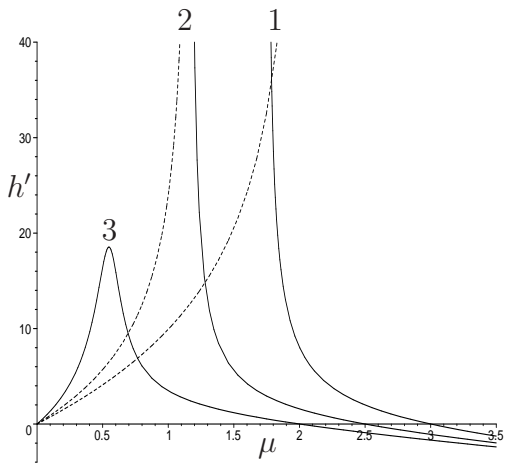}
\rput(-4,5.4){\large $1$}\rput(-5,4.2){\large $2$}
\rput(-6,3.3){\large $3$} \rput(-6,-3.2){\large $3$}
\rput(-5,-0.4){\large $2$}
 \rput(-3.8,-0.4){\large $1$} \rput(-7.2,3.4){\large $h$}\rput(-7.2,-2.6){\large $h'$}
 \rput(-3.9,0.4){\large $\mu$}\rput(-3.9,-6.1){\large $\mu$}
\caption{The top plot is $h$ vs $\mu$ for three different
solutions, all with $\phi=0$. The bottom plot shows $h'$ vs $\mu$
for the same solutions. ($1$):$\gamma=1/12-0.01,~\sigma=-3$.
($2$):$\gamma=1/12,~\sigma=-2.5$.
($3$):$\gamma=1/12+0.01,~\sigma=-2$. Different brane tensions are
used purely for clarity.} \label{mhg12}
\end{figure}
In Fig.~\ref{mhg12} we show results for $h$ and $h'$ for three
values of $\gamma$, $\gamma=1/12-0.01,~1/12,~1/12+0.01$. In the top
plot we see that the solution denoted (1) has GBIG1-3 present as
$\gamma<1/12$, we have two real and different values for $h_{\rm i}$
and $h_{\rm e}$. This solution in the bottom plot has three parts
(GBIG1 has negative values not shown in Fig.~\ref{mhg12}), GBIG2 is
the dotted solution and the solid is GBIG3. As ($1$) approaches
$h_{\rm i,e}$ $|h'|\rightarrow\infty$. Solution (2) has
$\gamma=1/12$, in the top plot we see that GBIG1 now no longer
exists as $h_{\rm i}=h_{\rm e}$. In the bottom plot GBIG3 (solid)
and GBIG2 (dotted) solutions both go to $h'=\infty$ at the point
$h_{\rm i}=h_{\rm e}$. Solution (3) has $\gamma>1/12$. Now $h_{\rm
i,e}$ does not represent $d\mu/dh=0$, therefore $h'$ stays well
behaved throughout the evolution (bottom plot). GBIG3 moves
seamlessly onto GBIG2 making a single solution. As GBIG2
super-accelerates this new solution will show late time phantom like
behaviour ($h'>0\rightarrow w<-1$).

Solutions that lie along the $\sigma_l$ line in Fig.~\ref{gs0} are
those considered in \cite{Brown:2005ug}. Regions I and II in
Fig.~\ref{gs0} extend up to $\gamma_{\rm M}=\infty$. Solutions in
each region of Fig.~\ref{gs0} have the following properties:

\begin{itemize}
\item $I:~\sigma_{\rm e}(\gamma_m)<\sigma<0,~\gamma>\gamma_{\rm i}$ and
$\sigma<\sigma_{\rm e}(\gamma_m),~\gamma>\gamma_{\rm m}$. GBIG1-2 do
not exist. GBIG3 loiters at $\mu_l$ before evolving to a vacuum de
Sitter universe.

\item $II:~\sigma>0,~\gamma>\gamma_{\rm i}$. GBIG1-2 do not exist.
GBIG3 evolves to a vacuum de Sitter universe.

\item $III:~\sigma\leq\sigma_{\rm e},~\gamma\leq\gamma_{\rm m}$. GBIG1 will evolve to
 ($\mu_{\rm e},h_{\rm e}$). GBIG2 evolves to ($0,h_{\infty}$). GBIG3 loiters at
$\mu_l$ before evolving towards ($\mu_{\rm e},h_{\rm e}$). When
$\sigma=\sigma_{\rm e}$ GBIG1 and 3 evolve to ($0,h_{\rm e}$). When
$\gamma=\gamma_{\rm m}$ GBIG1 ceases to exist ($h_{\rm i}=h_{\rm
e}$).

\item $IV:~\sigma_{\rm e}<\sigma<0,~\gamma\leq\gamma_{\rm i}$.
GBIG1-2 both evolve to vacuum de Sitter states. GBIG3 loiters before
evolving to a vacuum de Sitter universe. Each vacuum de Sitter state
has a different value of $h_{\infty}$. When $\gamma=\gamma_{\rm i}$
GBIG1-2 live at ($0,h_{\rm i}$).

\item $V:~\sigma\geq0,~\gamma\leq\gamma_{\rm i}$. GBIG1-3 all
end in vacuum de Sitter states. With $\gamma=\gamma_{\rm i}$ GBIG1-2
live at ($0,h_{\rm i}$). When $\sigma=0$ GBIG3 ends in a Minkowski
state.
\end{itemize}

\begin{figure}
\includegraphics[height=3in,width=2.75in,angle=270]{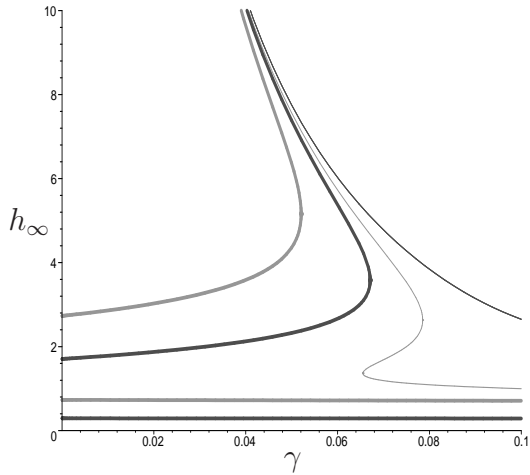}
\rput(-7.4,-3.5){\large $h_{\infty}$}\rput(-3.9,-6.7){\large
$\gamma$}
 \caption{$h_{\infty}$ for solutions in a Minkowski ($\phi=0$) bulk.
 The thin-dark line has $\sigma=-2$, thin-light line has $\sigma=-1.2$,
 thick-dark lines have $\sigma=-1/2$ and the thick-light lines have $\sigma=2$.}
\label{hinp0}
\end{figure}
In Fig.~\ref{hinp0} we can see how solutions in each of the regions
mentioned above affect $h_{\infty}$. The thin-dark line
($\sigma=-2$) lies in III and I in Fig.~\ref{gs0}. Only GBIG2 exists
for $0\leq\gamma<1/12$. For $\gamma>1/12$ we've seen that GBIG3 and
GBIG2 connect to make a single solution, which is why the line is
continuous through $\gamma=1/12$. The thin-light line
($\sigma=-1.2$) has an interesting feature due to the solutions
lying on a line that cuts through both region III and IV as well as
I. Both the thick lines have GBIG1-3 ending at $h_{\infty}$. The
thick-dark solution loiters before evolving to $h_{\infty}$, but
this has no effect upon the results in Fig.~\ref{hinp0}.

\subsection{AdS bulk ($\phi\neq0$) with brane tension}

When $\phi=0$, we have $\Lambda_5\neq0$, with one exception: the
$\phi_2$ solution with $\chi=0$, i.e. $\phi_2=-4/3\gamma$, has
$\Lambda_5=0$, but the bulk is AdS, Ref.~\cite{Brown:2005ug}. For
$\chi>0$, $\Lambda_5\neq0$. Thus in all cases, $\phi\neq0$ implies
an AdS bulk.

\begin{figure}
\includegraphics[height=3in,width=2.75in,angle=270]{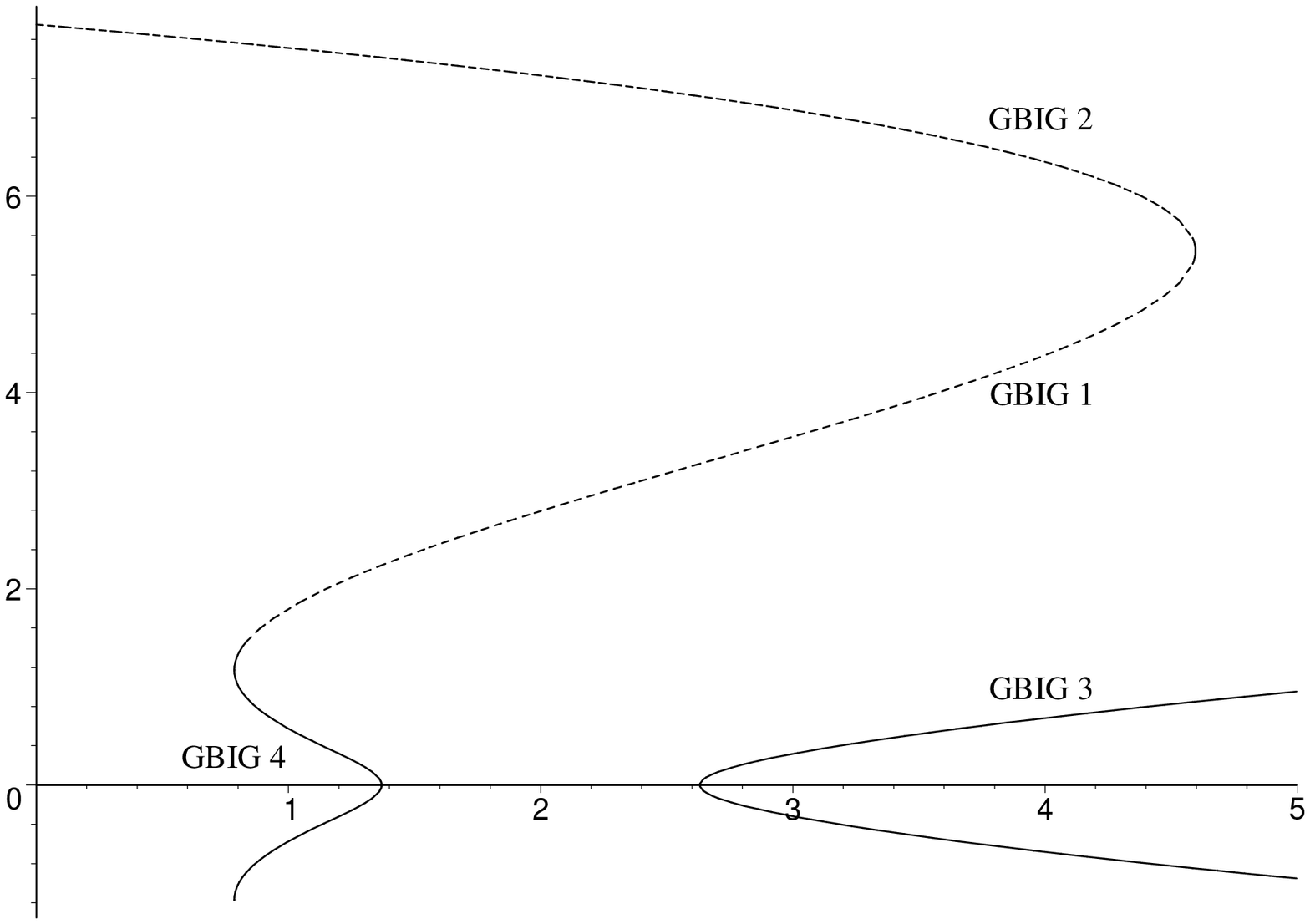}
\rput(-0.8,-2.1){\large$\mu_{\rm i},h_{\rm i}$}
\rput(-5,-5.3){\large$\mu_{\rm b}$} \rput(-4,-5.3){\large$\mu_{\rm
c}$}
 \rput(-6.4,-4.3){\large
$\mu_{\rm e},h_{\rm e}$} \rput(-7.3,-0.8){\large$h_{\infty}$}
\rput(-7.4,-3){\large $h$}\rput(-4.5,-5.9){\large $\mu$}
\rput{350}(-3.5,-1.2){\large $\leftarrow$}
\rput{20}(-3.5,-3.55){\large $\leftarrow$}
\rput{11}(-1.2,-4.8){\large $\leftarrow$}
\rput{325}(-5.6,-5.1){\large$\leftarrow$}
\rput{345}(-1.2,-6.2){\large $\rightarrow$}
\rput{30}(-5.6,-5.9){\large$\rightarrow$} \caption{Solutions of
the Friedmann equation ($h$ vs $\mu$) with negative brane tension
($\sigma=-2$) in an AdS bulk ($\phi=-0.1$) with $\gamma=1/20$. The
curves are independent of the equation of state $w$. The arrows
indicate the direction of proper time on the brane.}
\label{Friedpl1ds}
\end{figure}

When we allow the bulk to be warped ($\phi\neq0$) we
open up another possible solution, denoted GBIG4. There is a
maximum value of $\phi$ for which GBIG4 can exist as we shall see
later. The nature of GBIG3 is also changed. These solutions, for a
negative brane tension, can be seen in Fig.~\ref{Friedpl1ds}.
Brane tension affects the solutions in the same manner as in
$\phi=0$ case.
The qualitative effect of the warped bulk is to make GBIG3 collapse
and to introduce the new bouncing branch GBIG4. This is due to $h=0$
now giving two solutions:

\begin{equation}\label{mucP1}
\mu_{\rm c,b}=\pm\sqrt{-\phi}\left(2+\gamma\phi\right)-\sigma,
\end{equation}
where the plus sign is for the GBIG3 collapse ($\mu_{\rm c}$) and
the minus for the GBIG4 bounce ($\mu_{\rm b}$). Effectively the
loitering point in the Minkowski bulk is split into the max/min
densities of the bouncing/collapsing cosmologies of GBIG3-4. The
other points in Fig.~\ref{Friedpl1ds} are modified by the warped
bulk; they are now given by:

\begin{equation}\label{muim}
\mu_{\rm
i,e}=\frac{1-18\gamma+27\gamma^2\phi\pm(1-12\gamma-18\gamma^2\phi)^{3/2}}{54\gamma^2}
-\sigma,
\end{equation}
with the plus sign for $\mu_{\rm i}$ and the negative sign for
$\mu_{\rm e}$. The Hubble rates $h_{\rm i,e}$ are given by:

\begin{equation}\label{Hiem}
h_{\rm
i,e}=\frac{\sqrt{2}}{6\gamma}\sqrt{1-6\gamma+9\gamma^2\phi\pm\sqrt{1-12\gamma-18\gamma^2\phi}},
\end{equation}
with the same sign convention. $h_{\infty}$ is given by a
similarly modified equation:

\begin{eqnarray}\label{Phi1hinm}
h_{\infty}^6&+&\frac{(8\gamma-1)}{4\gamma^2}h_{\infty}^4+\frac{[4+2\sigma-
\phi\gamma(4+3\phi\gamma)]}{4\gamma^2}h_{\infty}^2\nonumber\\
&-&\frac{\phi(2+\phi\gamma)^2+\sigma^2}{4\gamma^2}=0.
\end{eqnarray}

\begin{figure}
\includegraphics[height=3in,width=2.75in,angle=270]{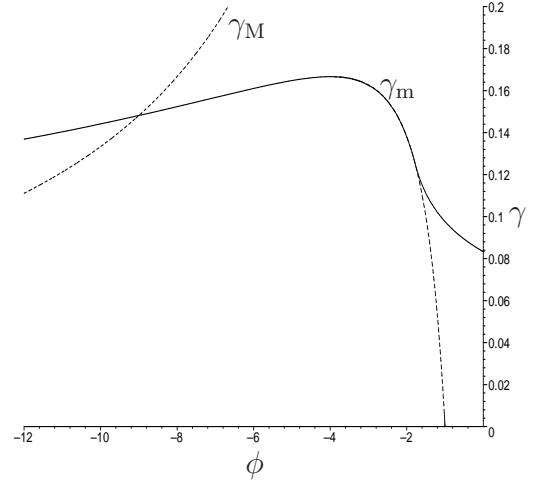}
\rput(-3.9,-6.8){\large
$\phi$}\rput(-0.4,-3.5){\large $\gamma$} \rput(-4,-1){\large
$\gamma_{\rm M}$} \rput(-2,-1.8){\large $\gamma_{\rm m}$}
 \caption{In order for $h_{\rm i}$ to be real $\gamma$
must lie beneath the solid line ($\gamma_{\rm m}$). The region to
the right of the vertical dotted line is where GBIG4 ($h_{\rm
e}>0$) is allowed (except where $\phi=0$). }\label{gpbound}
\end{figure}

With a warped bulk the constraint from $h_{\rm i}$ ($\gamma\leq1/12$
in the Minkowski case) is modified. In our equation for $h_{\rm i}$
we effectively have two bounds from the two square root terms. In
the $\phi=0$ case only the inner term is of any consequence (giving
rise to the bound $\gamma\leq1/12$). When $\phi\neq0$ there are two
bounds which are applicable in different regimes. If we consider the
bound from the inner square root we get:

\begin{equation}\label{gbound}
\gamma\leq\frac{\sqrt{4+2\phi}-2}{6\phi}.
\end{equation}
This is valid for $\phi\geq-2$. Considering the outer square root
term we get the bound:

\begin{equation}\label{gbound2}
\gamma\leq\frac{2\left(1-\sqrt{-\phi}\right)}{3\phi}.
\end{equation}
This bound becomes negative when $\phi>-1$ which is unallowed due to
our string constraint. The bound on $\gamma$ is given by the lower
of the two constraints when $-2<\phi<-1$ i.e. when in the range
where both exists.  Therefore the bound on $\gamma$ for $h_{\rm i}$
to be real is given by (see Fig.~\ref{gpbound}):

\begin{equation}\label{gboundT}
\gamma\leq\left\{ \begin{array}{lc}
 \frac{\sqrt{4+2\phi}-2}{6\phi}  &-\frac{16}{9}\leq\phi\leq0\\\\
 \frac{2\left(1-\sqrt{-\phi}\right)}{3\phi}  &\phi\leq-\frac{16}{9}\\
  \end{array}
 \right.
\end{equation}
For a particular value of $\phi$ the maximum value of $\gamma$
allowed by this constraint is denoted $\gamma_{\rm m}$. When
$\phi<-9$ the constraint in Eq.~(\ref{acon2}) is tighter than that
in Eq.~(\ref{gbound2}) ($\gamma_{\rm M}<\gamma_{\rm m}$). This
means that for $\phi\leq-9$, $h_{\rm i}$ is always real for
allowed values of $\gamma$.

There is a bound for GBIG4 to exist, found by considering $h_{\rm
e }=0$. This bound is given by:

\begin{equation}\label{chib}
\phi>-\frac{2}{9\gamma^2}\left\{1-3\gamma-\sqrt{1-6\gamma}\right\}.
\end{equation}
This is a solution to the quadratic obtained from $h_{\rm e}=0$, the
other root of the quadratic does not obey the constraint $\phi\leq
-4/3\gamma$ so is ignored. Using our constraints we can split the
$\gamma,~\phi$ plane into three sections, see Fig.~\ref{gpbound}.
The solid line is constructed from the bounds in
Eq.~(\ref{gboundT}). For $h_{\rm i}$ to be real we must choose
values below this line. The area to the right of the vertical dotted
line (but excluding $\phi=0$) allows GBIG4. Points to left of this
line have GBIG1 collapsing after a minimum energy density ($\mu_{\rm
b}$) is reached. The dotted curve on the left comes from initial
constraint $\gamma\leq -4/3\phi$.

In order to consider the $\sigma,~\gamma$ plane as we did in the
$\phi=0$ Minkowski case, we need the modified $\sigma_{\rm i,e}$
equations:

\begin{equation}\label{silm}
\sigma_{\rm
i,e}=\frac{1-18\gamma+27\gamma^2\phi\pm(1-12\gamma-18\gamma^2\phi)^{3/2}}{54\gamma^2}.
\end{equation}
We now have two more formulas for the collapse density of GBIG3
and the bounce density of GBIG4:

\begin{equation}\label{sicP1m}
\sigma_{\rm c,b}=\pm\sqrt{-\phi}\left(2+\gamma\phi\right),
\end{equation}
with the plus sign corresponding to the collapse and the minus to
the bounce.

We shall consider the $\sigma,~\gamma$ plane for three different
values of $\phi$ corresponding to three distinct regions in
Fig.~\ref{gpbound}. We shall first consider $\phi=-1/2$.

\subsubsection{Typical example $\phi=-1/2$}

\begin{figure}
\includegraphics[height=3in,width=2.75in,angle=270]{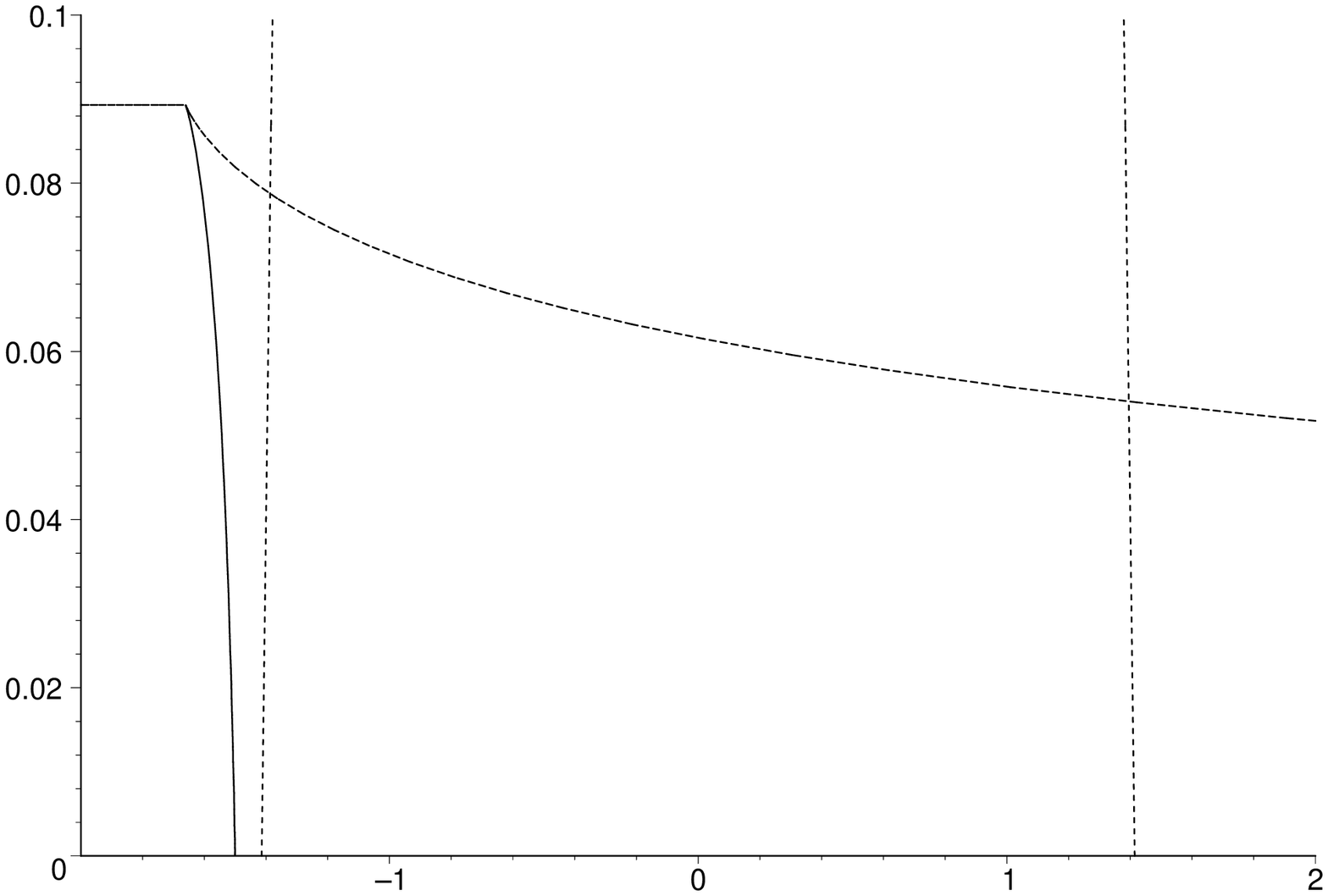}
\rput(-7.4,-3.6){\large
$\gamma$}\rput(-3.9,-6.7){\large $\sigma$}
\psline{->}(-7.9,-2.8)(-6.3,-3)\rput(-6.5,-1.25){\large
I}\rput(-4,-1.25){\large II} \rput(-1,-1.25){\large III}
\rput(-6.6,-4.5){\large IV} \rput(-6.2,-2.4){\small
V}\rput(-4,-4.5){\large VI}\rput(-1,-4.5){\large VII}
 \rput(-8,-2.5){\large $\sigma_{\rm
e}(\gamma)$} \rput(-1,-3.2){\large $\gamma_{\rm
i}(\sigma)$}\rput(-2.3,-6){\large $\sigma_{\rm
c}(\gamma)$}\rput(-5.5,-6){\large $\sigma_{\rm
b}(\gamma)$}\rput(-7.3,-1.6){\large $\gamma_{\rm m}$} \caption{The
$(\sigma,\gamma)$ plane for solutions in a AdS ($\phi=-1/2$) bulk.
The short dotted horizontal line is the maximum value of $\gamma$
as obtained from Eq.~(\ref{acon2}).} \label{gs12}
\end{figure}

If we take $\phi=-1/2$, we are in the region where GBIG4 is
allowed. The $\sigma,~\gamma$ plane can be seen in
Fig.~\ref{gs12}. The regions I, II and III in Fig.~\ref{gs12}
extend up to $\gamma_{\rm M}=8/3$.

In each region we have the following cosmologies.
\begin{itemize}
\item
$I:~\sigma_{\rm e}(\gamma_m)<\sigma<0,~\gamma>\gamma_{\rm i}$ and
$\sigma\leq\sigma_{\rm e}(\gamma_m),~\gamma>\gamma_{\rm m}$. GBIG1-2
do not exist. GBIG3 reaches a minimum energy density ($\mu_{\rm c}$)
and then evolves back to $\mu=\infty$. GBIG4 starts at
($0,-h_{\infty}$) then bounces at ($\mu_{\rm b},0$) before evolving
to ($0,+h_{\infty}$). When $\sigma=\sigma_{\rm b}$ GBIG4 exists as a
Minkowski universe.

\item $II:~\sigma_{\rm b}\leq\sigma<\sigma_{\rm c},~ \gamma>\gamma_{\rm
i}$. GBIG1-2 and 4 do not exist. GBIG3 reaches a minimum energy
density ($\mu_{\rm c}$) and then evolves back to $\mu=\infty$. When
$\sigma=\sigma_{\rm c}$ GBIG3 ends in a Minkowski universe.

\item $III:~\sigma>\sigma_{\rm c},~ \gamma>\gamma_{\rm
i}$.  GBIG1-2 and 4 do not exist. GBIG3 evolves to a vacuum de
Sitter universe.

\item $IV:~\sigma\leq\sigma_{\rm e}<0,~\gamma\leq\gamma_{\rm m}$. GBIG1
evolves to ($\mu_{\rm e},h_{\rm e}$). GBIG2 evolves to $h_{\infty}$.
 GBIG3 reaches a minimum energy density ($\mu_{\rm c}$) and then
evolves back to $\mu=\infty$. GBIG4 starts at ($\mu_{\rm e},-h_{\rm
e}$), bounces at ($\mu_{\rm b},0$) before evolving to ($\mu_{\rm
e},+h_{\rm e}$). When $\sigma=\sigma_{\rm e}$ GBIG1 and 3 evolve to
($0,h_{\rm e}$). GBIG4 starts at ($0,-h_{\rm e}$) and bounces before
evolving to ($0,h_{\rm e}$). When $\gamma=\gamma_{\rm m}$ GBIG1
ceases to exist ($h_{\rm i}=h_{\rm e}$).

\item $V:~\sigma_{\rm e}<\sigma\leq\sigma_{\rm b},~\gamma\leq\gamma_{\rm i}$.
 GBIG1-2 both end in vacuum de Sitter universes with different
values of $h_{\infty}$. GBIG3 reaches a minimum energy density
($\mu_{\rm c}$) and then evolves back to $\mu=\infty$. GBIG4 starts
at ($0,-h_{\infty}$) then bounces at ($\mu_{\rm b},0$) before
evolving back to ($0,+h_{\infty}$). When $\sigma=\sigma_{\rm b}$
GBIG4 exists as a Minkowski universe ($0,0$).

\item $VI:~\sigma_{\rm b}<\sigma\leq\sigma_{\rm c},~\gamma\leq\gamma_{\rm
i}$. GBIG1-2 both end in vacuum de Sitter universes with different
values of $h_{\infty}$. GBIG3 reaches a minimum energy density
($\mu_{\rm c}$) and then evolves back to $\mu=\infty$. GBIG4 does
not exist. When $\sigma=\sigma_{\rm c}$ GBIG3 ends in a Minkowski
universe. When $\gamma=\gamma_{\rm i}$ GBIG1-2 live at ($0,h_{\rm
i}$).

\item $VII:~\sigma>\sigma_{\rm c},~\gamma\leq\gamma_{\rm i}$. GBIG1-3 all
end in vacuum de Sitter universes with different values of
$h_{\infty}$. GBIG4 does not exist. When $\gamma=\gamma_{\rm i}$
GBIG1-2 live at ($0,h_{\rm i}$).
\end{itemize}

\begin{figure}
\includegraphics[height=3in,width=2.75in,angle=270]{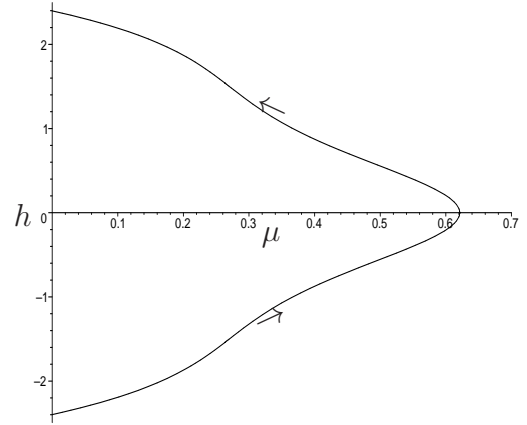}
\includegraphics[height=3in,width=2.75in,angle=270]{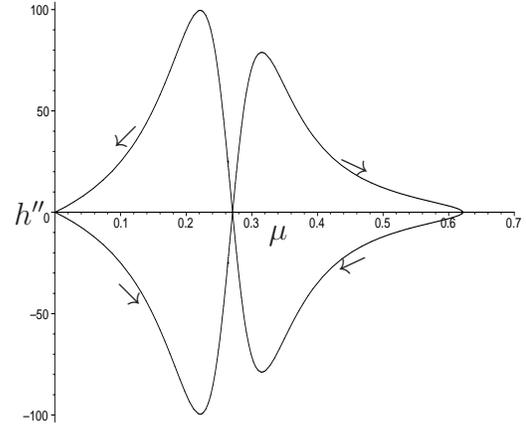}
\rput(-4,3.2){\large
$\mu$}\rput(-4,-3.8){\large $\mu$} \rput{335}(-4,4.9){\large
$\leftarrow$} \rput{25}(-4,2.1){\large $\rightarrow$}
\rput{45}(-6,-2.5){\large $\leftarrow$} \rput{315}(-6,-4.6){\large
$\rightarrow$}\rput{25}(-3,-4.2){\large $\leftarrow$}
\rput{335}(-3,-2.9){\large $\rightarrow$}
 \rput(-7.3,-3.5){\large $h''$}\rput(-7.3,3.5){\large $ h$} \caption{The GBIG4 solution in
region I for $\phi=-1/2,\gamma=0.1>\gamma_{\rm m}$ and
$\sigma=-2$. The bottom plot is $ h''$ vs $\mu$ for this solution.
This shows the differences between this solution and the similar
case when $\phi=-4$, where GBIG4 is no longer allowed. Arrows
denote proper time.} \label{uh12I}
\end{figure}

In region I there is a GBIG4 solution, which is modified in the same
way as the GBIG3 solution in the Minkowski bulk. As $h_{\rm i,e}$
are no longer valid in region I GBIG4 joins onto GBIG2, see
Fig.~\ref{uh12I}.
The bottom plot in this figure shows $h''$, in order to clearly
distinguish this solution from the one in the $\phi=-4$ case where
GBIG4 is no longer allowed.

\begin{figure}
\includegraphics[height=3in,width=2.75in,angle=270]{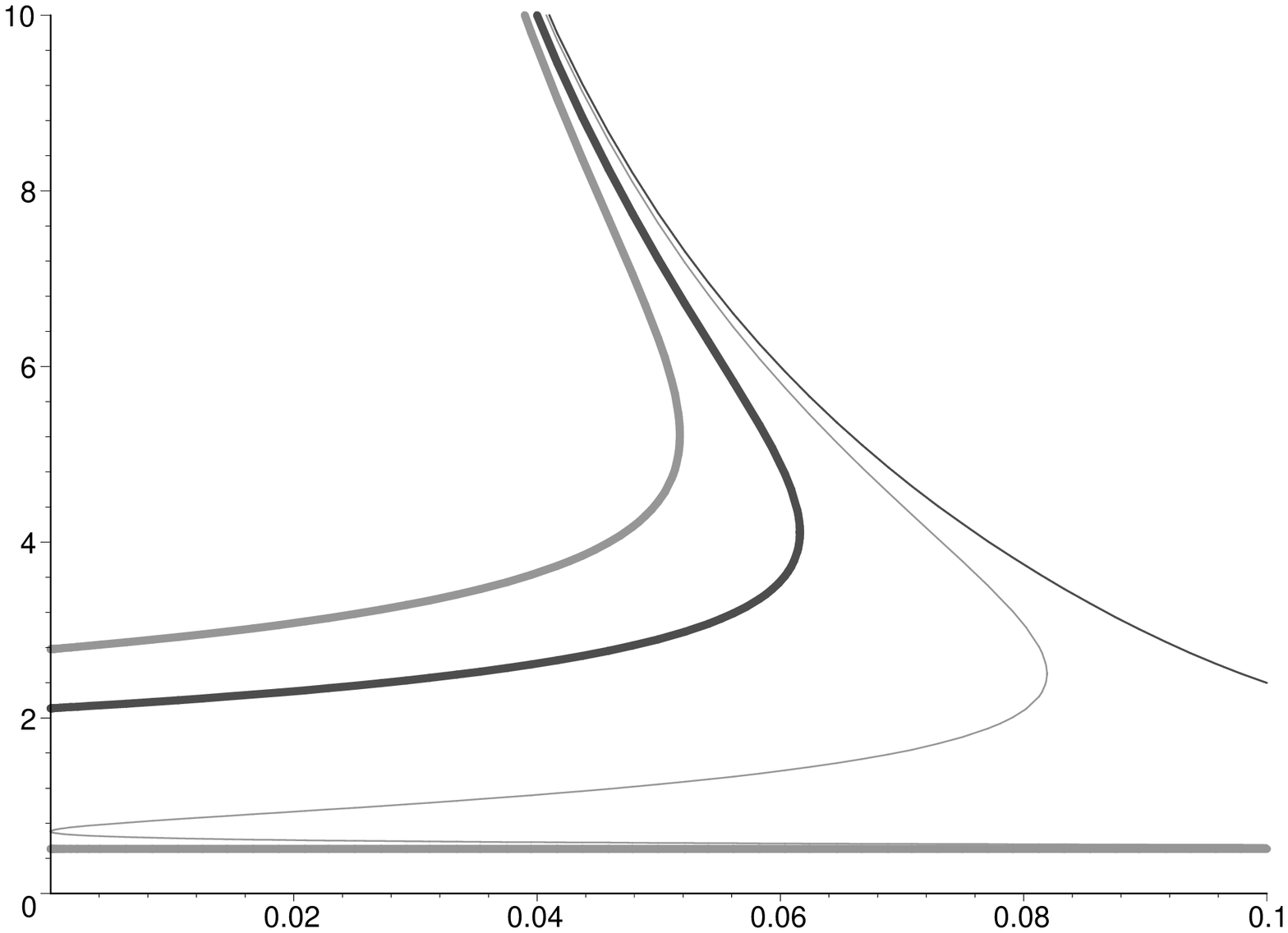}
\rput(-7.4,-3.5){\large $h_{\infty}$}\rput(-3.9,-6.7){\large
$\gamma$}
 \caption{$h_{\infty}$ for solutions in a AdS ($\phi=-1/2$) bulk.
 The thin-dark line has $\sigma=-2$, thin-light lines have $\sigma=-1.5$,
 thick-dark line has $\sigma=0$ and the thick-light lines have $\sigma=2$.}
\label{hinp12}
\end{figure}

In Fig.~\ref{hinp12} are the results for $h_{\infty}$ with
$\phi=-1/2$. The thin-dark line ($\sigma=-2$) lies in regions I and
IV. Therefore we have only one solution for $h_{\infty}$, which
corresponds to GBIG2 for $\gamma\leq\gamma_{\rm m}$. For
$\gamma_{\rm m}<\gamma<\gamma_{\rm M}$ this root now corresponds to
the end point for GBIG3 as in the $\phi=0$ case. The thin-light
lines ($\sigma=-1.5$) lie in regions I and V. There are the two
GBIG1-2 solutions and the GBIG4 solution which converges with the
light-thick line at the bottom. The thick-dark line ($\sigma=0$)
lies in regions II and VI, so only has GBIG1-2 present. The
thick-light lines ($\sigma=2$) lie in regions III and VII, so has
GBIG1-2 and the GBIG3 solution (the horizontal line).

\subsubsection{Typical example: $\phi=-4$}

\begin{figure}
\includegraphics[height=3in,width=2.75in,angle=270]{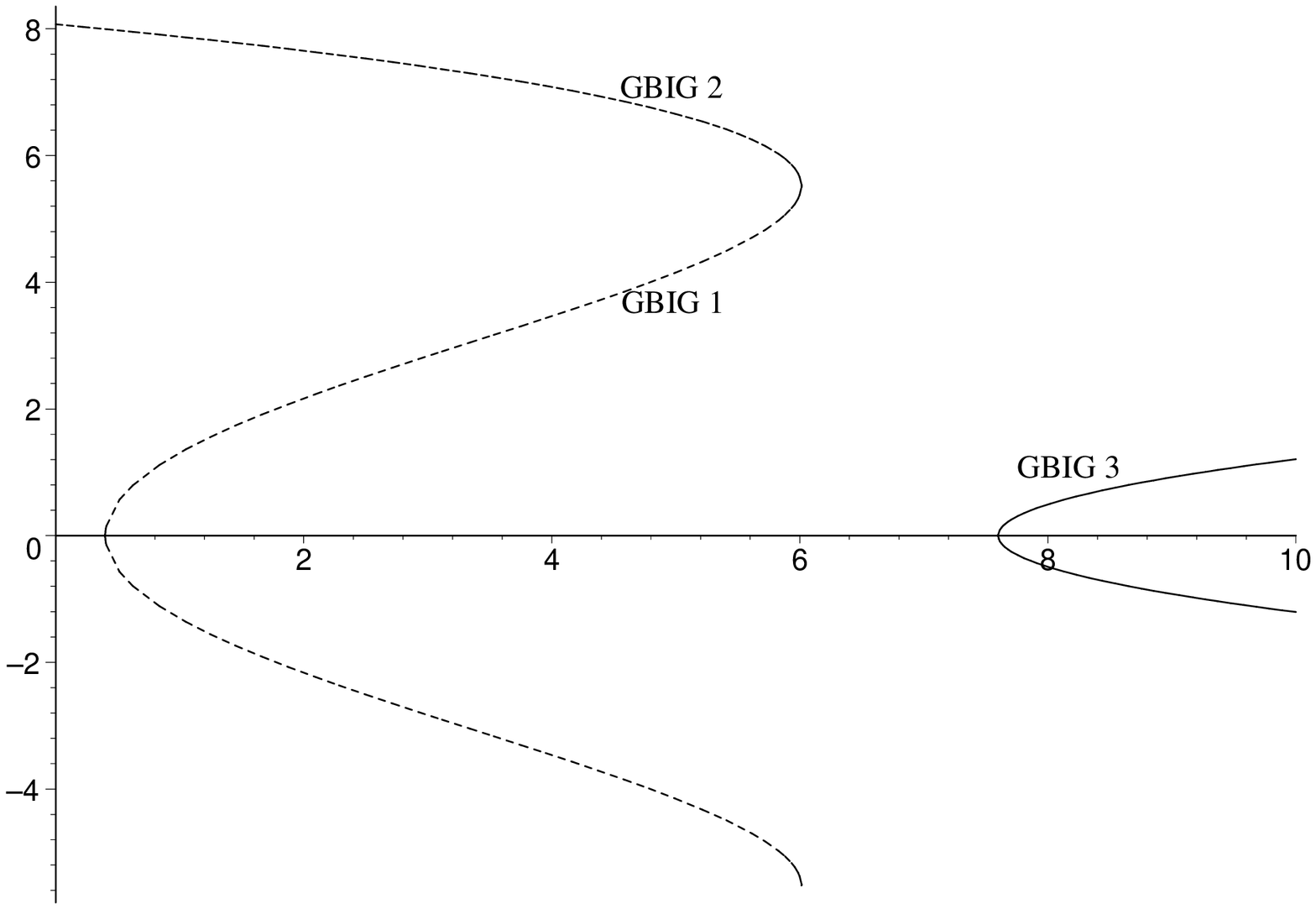}
\rput(-2.7,-6.2){\large$\mu_{\rm i},-h_{\rm
i}$}\rput(-2.7,-1.8){\large$\mu_{\rm i},h_{\rm i}$}
\rput(-6.7,-3.4){\large$\mu_{\rm b}$}
\rput(-2.5,-3.8){\large$\mu_{\rm c}$}
  \rput(-7.5,-0.8){\large$h_{\infty}$}
\rput(-7.4,-3){\large $h$}\rput(-4,-4.3){\large $\mu$}
\rput{350}(-5,-1){\large $\leftarrow$} \rput{20}(-5,-3){\large
$\leftarrow$} \rput{11}(-1,-3.4){\large $\leftarrow$}
\rput{345}(-1,-4.6){\large $\rightarrow$}
\rput{330}(-5,-5){\large$\rightarrow$} \caption{Solutions of the
Friedmann equation ($h$ vs $\mu$) with negative brane tension
($\sigma=-4$) in an AdS bulk ($\phi=-4$) with $\gamma=1/20$. For
this value of $\phi$ GBIG4 no longer exists and GBIG1 can collapse.
The curves are independent of the equation of state $w$. The arrows
indicate the direction of proper time on the brane.} \label{Phi12}
\end{figure}

\begin{figure}
\includegraphics[height=3in,width=2.75in,angle=270]{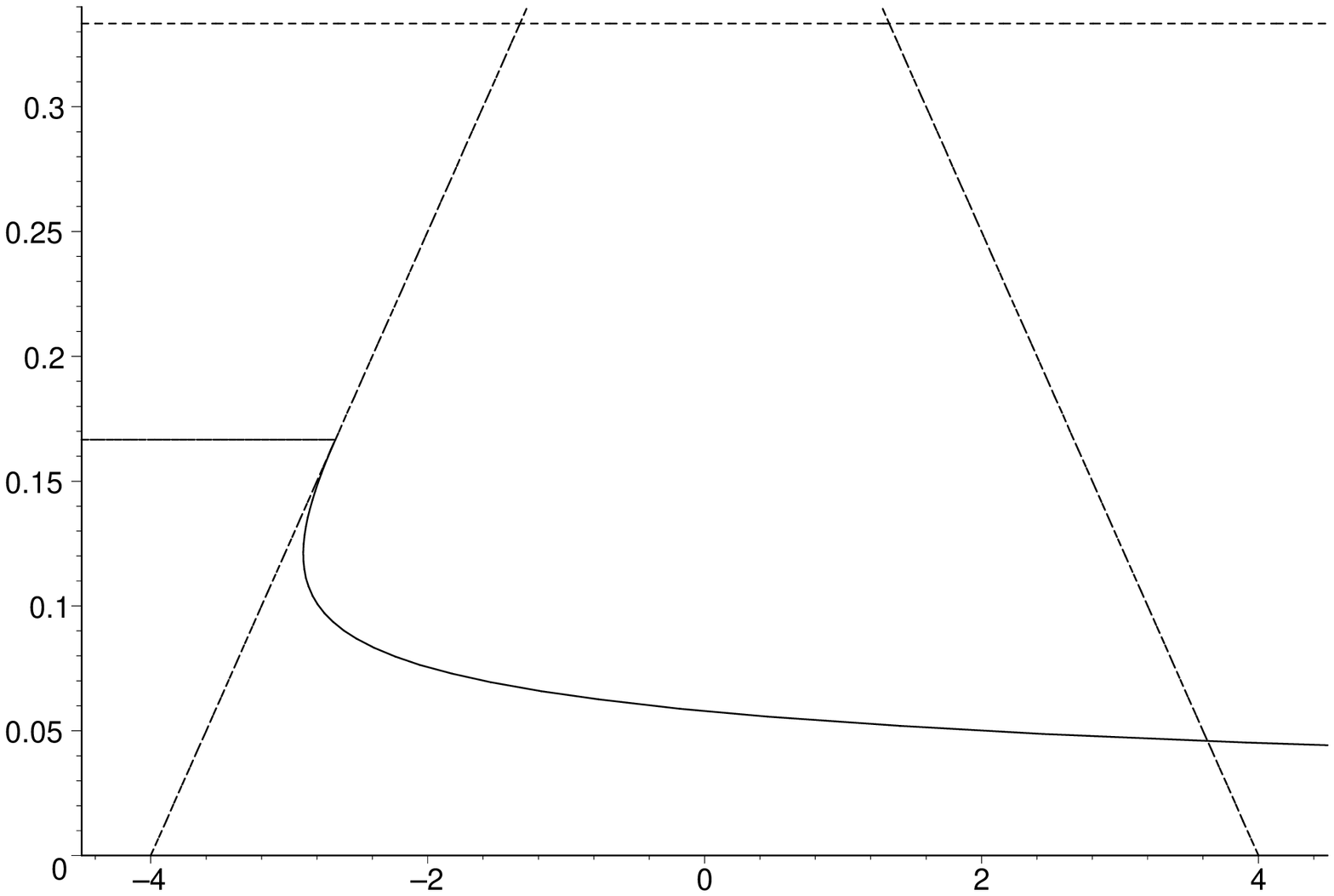}
\rput(-7.4,-3.6){\large
$\gamma$}\rput(-3.9,-6.7){\large $\sigma$} \rput(-6.2,-2.4){\large
I}\rput(-4,-2.4){\large II} \rput(-1,-2.4){\large III}
\rput(-6.6,-4.5){\large IV} \rput(-4,-6){\large
V}\rput(-1,-6){\large VI}
 \rput(-4.8,-4.9){\large $\gamma_{\rm i}(\sigma)$}\rput(-1.2,-4.6){\large
$\sigma_{\rm c}(\gamma)$}\rput(-5.9,-5.8){\large $\sigma_{\rm
b}(\gamma)$}\rput(-6.5,-3.5){\large $\gamma_{\rm m}$}
\rput(-6.5,-0.9){\large $\gamma_{\rm M}$}\caption{The
$(\sigma,\gamma)$ plane for solutions in a AdS ($\phi=-4$) bulk. The
short dotted horizontal line is the maximum value of $\gamma$ as
obtained from Eq.~(\ref{gbound}). The top horizontal line is from
the initial bound in Eq.~(\ref{gpbound}).} \label{gs4}
\end{figure}

We now consider $\phi=-4$ which is in the region where GBIG4 no
longer exists (see Fig.~\ref{Phi12}), as $h_{\rm e}=0$.
Therefore $\mu_{\rm e}$ is not relevant and GBIG1 bounces at
$\mu_{\rm b}$. This means that the ($\sigma,~\phi$) plane,
Fig.~\ref{gs4}, is simpler.
The regions in Fig.~\ref{gs4} are:

\begin{itemize}
\item$I:~\sigma\leq\sigma_{\rm
b},~\gamma_{\rm m}<\gamma\leq\gamma_{\rm M}$. GBIG1 does not exist.
GBIG2 starts at ($0,-h_{\infty}$), collapses to ($\mu_{\rm b},0$)
and then expands back to ($0,h_{\infty}$). GBIG3 expands to
($\mu_{\rm c},0$) and then collapses. When $\sigma=\sigma_{\rm b}$
GBIG2 exists as a Minkowski universe.

\item $II:~\sigma_{\rm b}<\sigma\leq\sigma_{\rm
c},~\gamma_{\rm i}<\gamma\leq\gamma_{\rm M}$. GBIG1-2 do not exist.
GBIG3 expands to $\mu_{\rm c}$ and then collapses. When
$\sigma=\sigma_{\rm c}$ GBIG3 evolves to a Minkowski universe.

\item $III:~\sigma>\sigma_{\rm c},~\gamma_{\rm i}<\gamma\leq\gamma_{\rm M}$.
 GBIG1-2 do not exist. GBIG3 evolves to $h_{\infty}$.

\item $IV:~\sigma\leq\sigma_{\rm b},~\gamma\leq\gamma_{\rm m}$. GBIG1
evolves from $\mu_{\rm i}$ to $\mu_{\rm b}$ and then expands back to
$\mu_{\rm i}$. GBIG2 evolves to $h_{\infty}$. GBIG3 expands to
$\mu_{\rm c}$ and then collapses. When $\gamma=\gamma_{\rm m}$ GBIG1
ceases to exist, GBIG2 expands from ($\mu_{\rm b},0$). When
$\sigma=\sigma_{\rm b}$ GBIG1 evolves to a Minkowski universe.

\item $V:~\sigma_{\rm b}<\sigma\leq\sigma_{\rm
c},~\gamma\leq\gamma_{\rm i}$. GBIG1-2 evolve to $h_{\infty}$. GBIG3
expands to $\mu_{\rm c}$ and then collapses. When
$\gamma=\gamma_{\rm i}$ GBIG1-2 exist as the same de Sitter universe
with ($0,h_{\infty}$). When $\sigma=\sigma_{\rm c}$ GBIG3 ends as a
Minkowski universe.

\item $VI:~\sigma>\sigma_{\rm c},~\gamma\leq\gamma_{\rm i}$.
GBIG1-3 evolve to $h_{\infty}$. When $\gamma=\gamma_{\rm i}$ GBIG1-2
live at ($0,h_{\infty}$).
\end{itemize}

\begin{figure}
\includegraphics[height=3in,width=2.75in,angle=270]{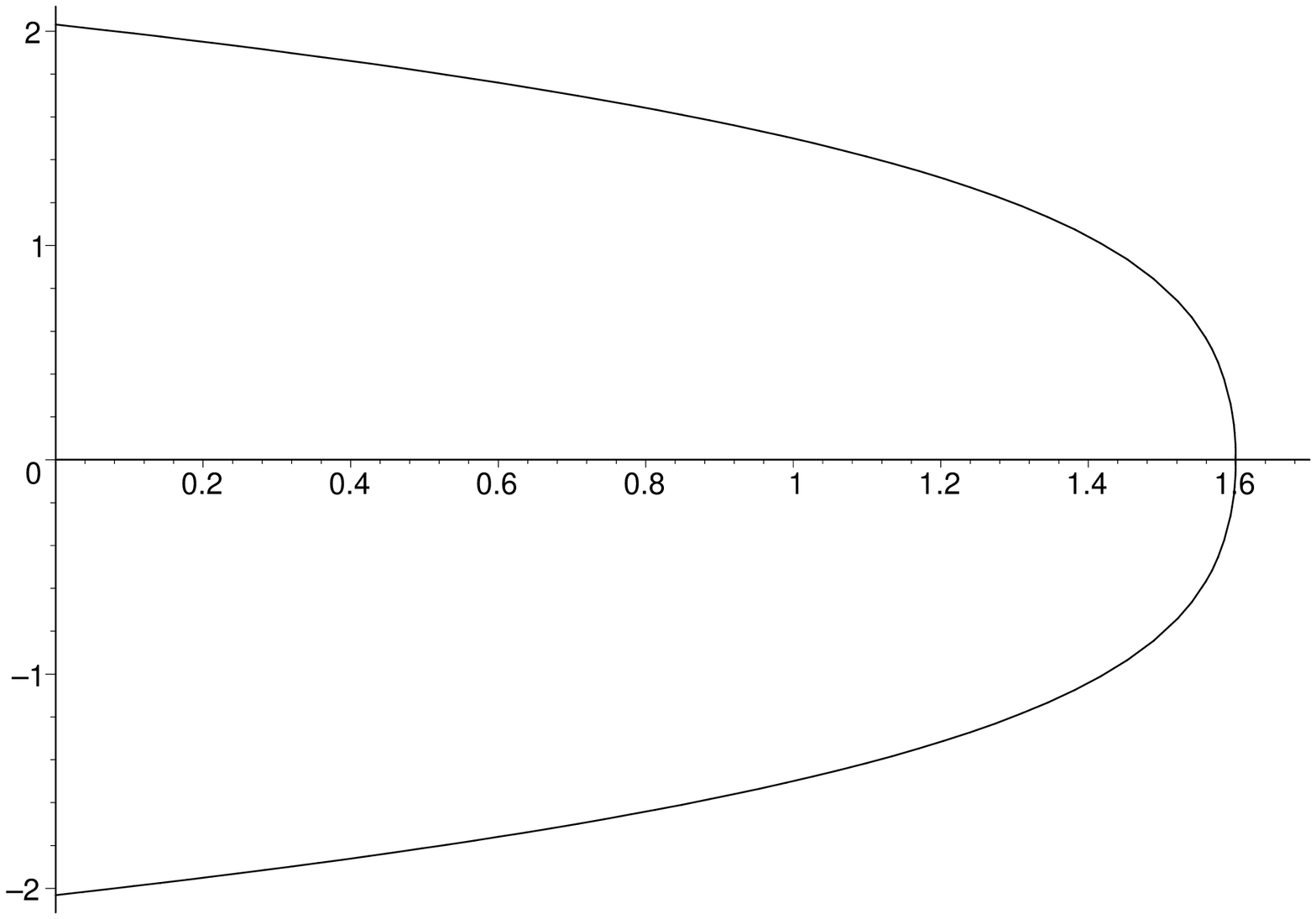}
\includegraphics[height=3in,width=2.75in,angle=270]{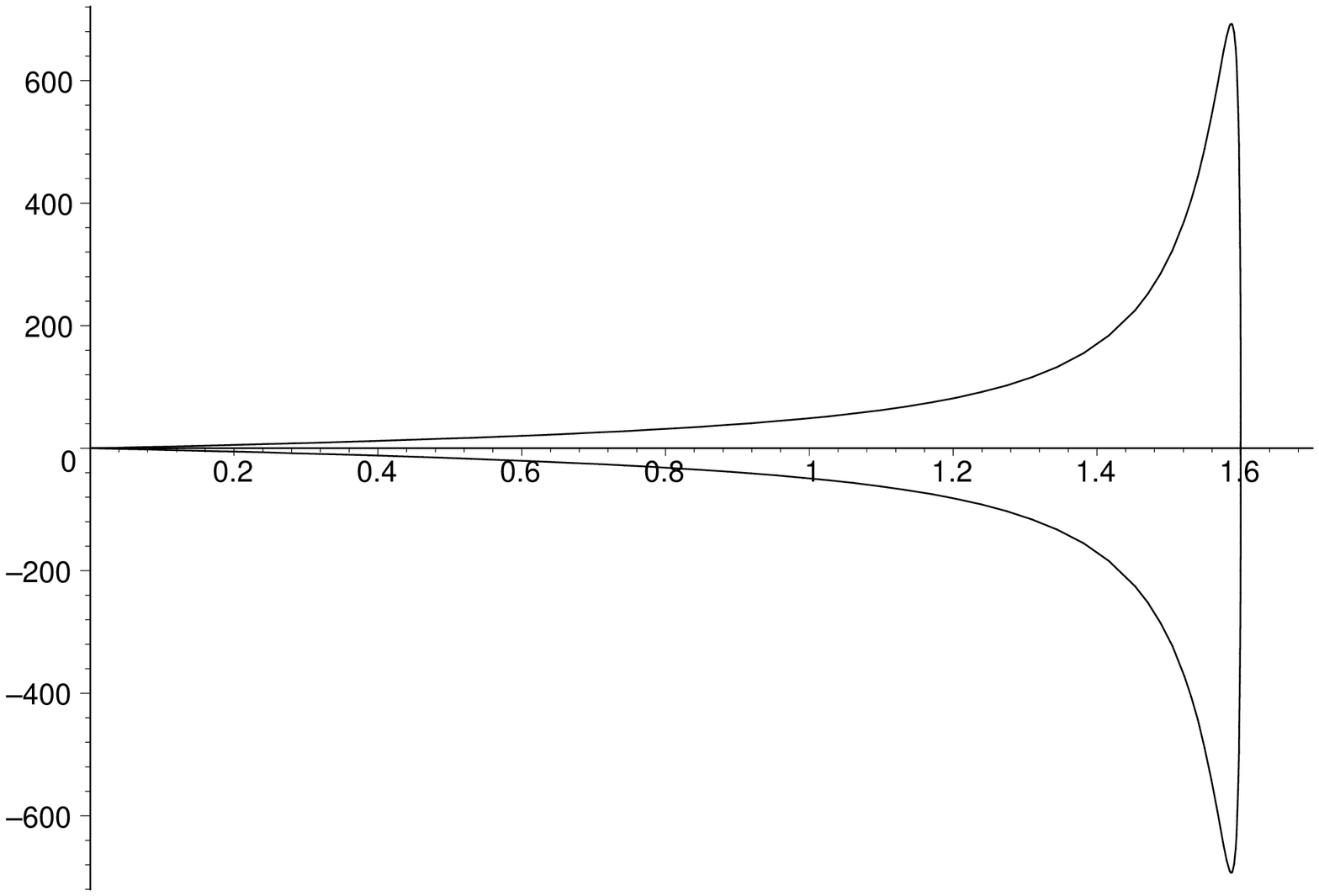}
\rput(-4,3){\large
$\mu$}\rput(-4,-4){\large $\mu$} \rput{345}(-4,5.8){\large
$\leftarrow$} \rput{15}(-4,1.2){\large $\rightarrow$}
\rput{45}(-2,-2.6){\large $\leftarrow$} \rput{315}(-2,-4.4){\large
$\rightarrow$} \rput(-7.3,-3.5){\large $h''$}\rput(-7.3,3.5){\large
$h$} \caption{The GBIG4 solution in region I for
$\phi=-4,\gamma=0.2>\gamma_{\rm m}$ and $\sigma=-4$. The bottom plot
is $h''$ vs $\mu$ for this solution. This shows the differences
between this solution and the case when $\phi=-1/2$. Arrows denote
proper time.} \label{uh4I}
\end{figure}

\begin{figure}
\includegraphics[height=3in,width=2.75in,angle=270]{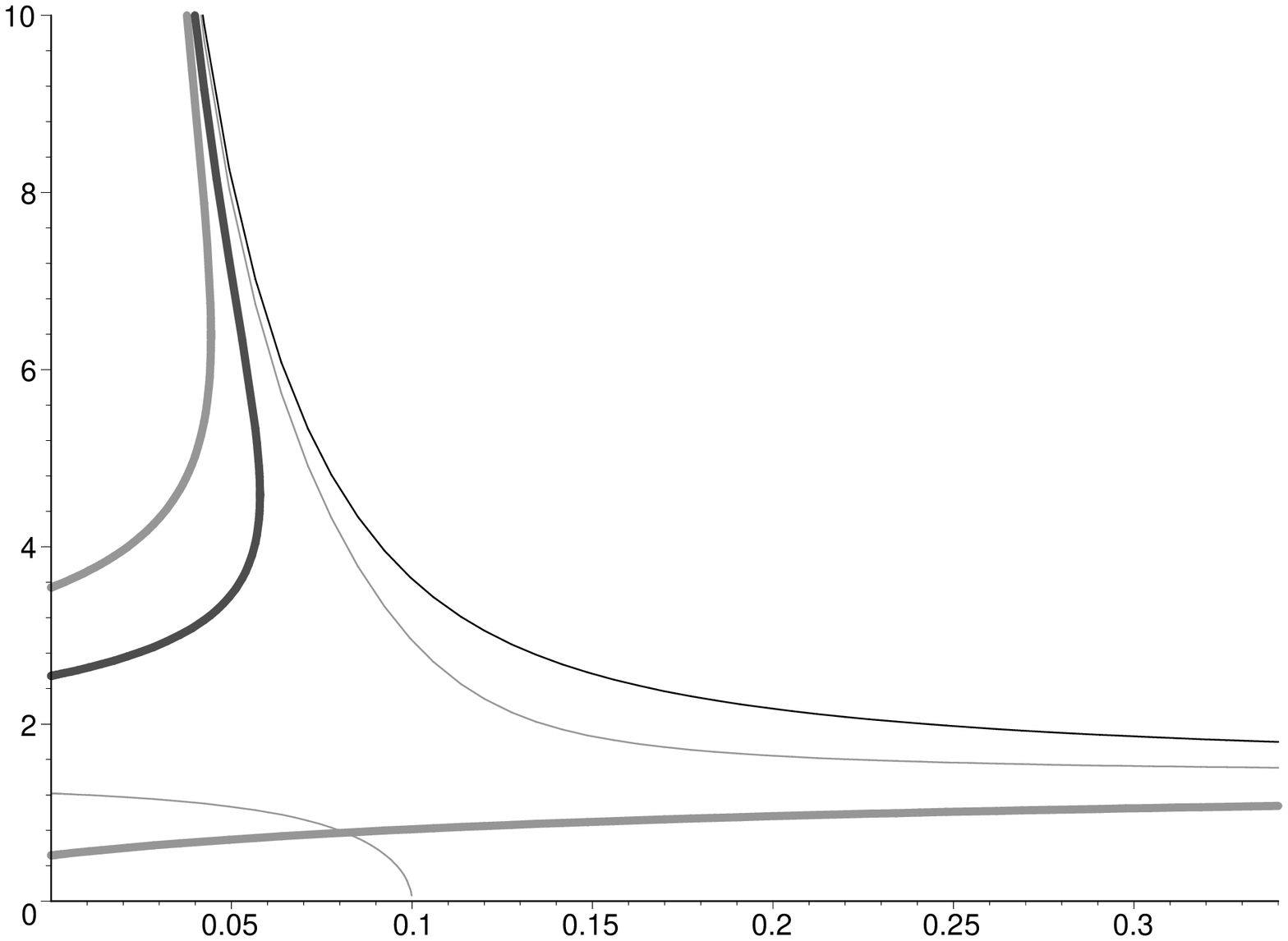}
\rput(-7.4,-3.5){\large $h_{\infty}$}\rput(-3.9,-6.7){\large
$\gamma$}
 \caption{$h_{\infty}$ for solutions in a AdS ($\phi=-4$) bulk.
 The thin-dark line has $\sigma=-4.4$, thin-light lines have $\sigma=-3.2$,
 thick-dark line has $\sigma=0$ and the thick-light lines have $\sigma=4.4$.}
\label{hinp4}
\end{figure}

In region I we again have a combined solution as GBIG1 has
vanished. As GBIG4 is not allowed ($h_{\rm e}=0$) when we take
$\gamma>\gamma_{\rm m}$, which causes $h_{\rm i}=0$, GBIG2 matches
up with its negative counterpart. We see the bouncing GBIG2
solution in Fig.~\ref{uh4I}. Note the nature of $h''$ is very
different to that of the GBIG2, 4 bounce in the $\phi=-1/2$ case.
In Fig.~\ref{hinp4} we present results for $h_{\infty}$ for
$\phi=-4$.

\subsubsection{Typical example: $\phi=-20$}

\begin{figure}
\includegraphics[height=3in,width=2.75in,angle=270]{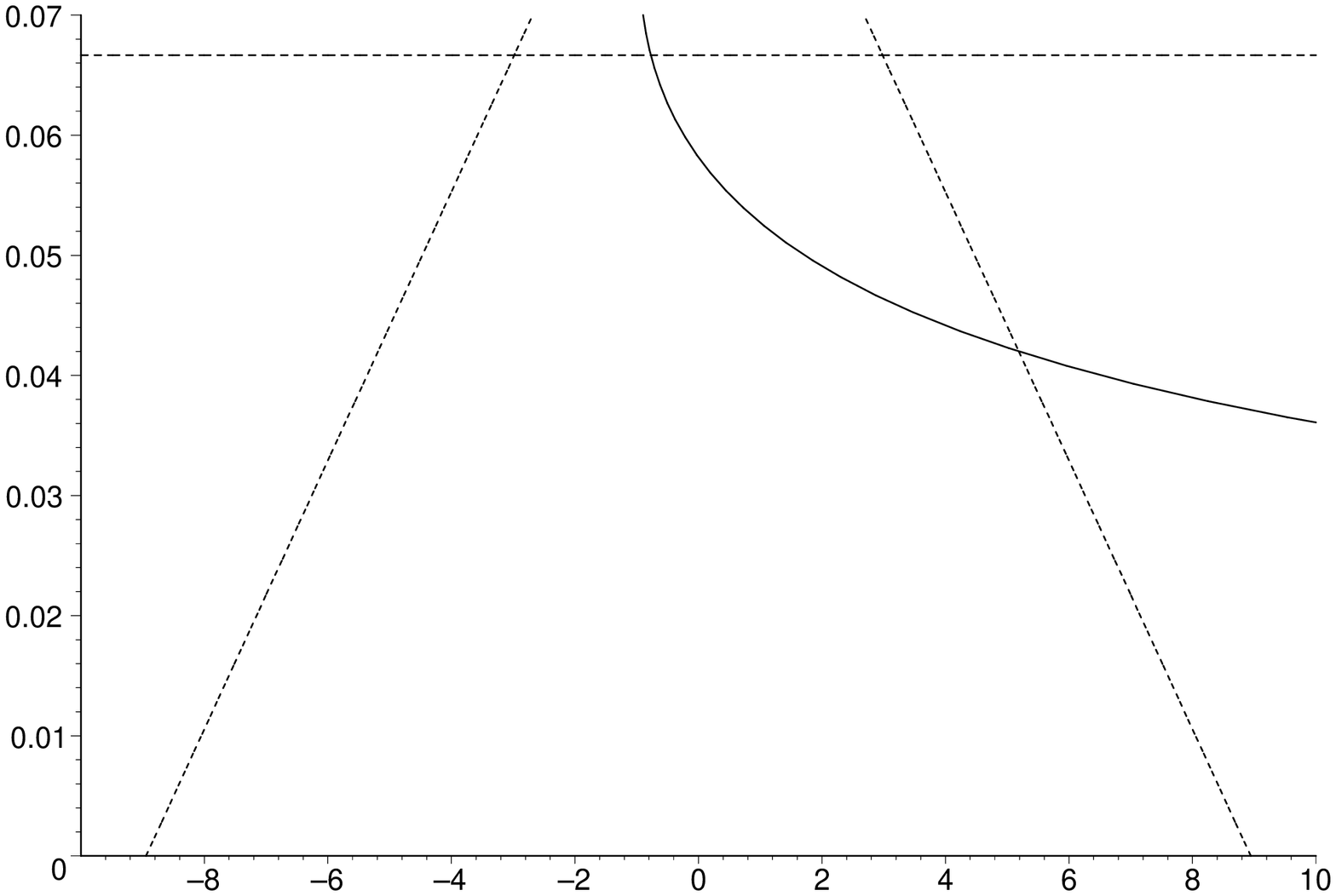}
\rput(-7.4,-3.7){\large
$\gamma$}\rput(-3.9,-6.7){\large $\sigma$}\rput(-6,-2){\large
I}\rput(-3.2,-2){\large II} \rput(-1.2,-2){\large III}
\rput(-4,-4.5){\large IV}\rput(-1.2,-4.5){\large V}
\rput(-1,-3.2){\large $\gamma_{\rm i}(\sigma)$}\rput(-1.8,-6){\large
$\sigma_{\rm c}(\gamma)$}\rput(-6,-6){\large $\sigma_{\rm
b}(\gamma)$}\rput(-6.5,-1){\large $\gamma_{\rm M}$} \caption{The
$(\sigma,\gamma)$ plane for solutions in a AdS ($\phi=-20$) bulk.
The top horizontal line is the bound from our initial bound in
Eq.~(\ref{gpbound}).} \label{gs20}
\end{figure}

Here we shall present results for $\phi=-20$ for completeness. This
value of $\phi$ lives in the region of Fig.~\ref{gpbound} where
$h_{\rm i}$ is always real. This means that the $\sigma,~\gamma$
plane is much simpler, Fig.~\ref{gs20}.

The regions in Fig.~\ref{gs20} are:

\begin{itemize}
\item$I:~\sigma\leq\sigma_{\rm
b},~\gamma\leq\gamma_{\rm M}$. GBIG1 evolves from $\mu_{\rm i}$ to
$\mu_{\rm b}$ and back. GBIG2 evolves to $h_{\infty}$. GBIG3 expands
to $\mu_{\rm c}$ and then collapses. When $\sigma=\sigma_{\rm b}$
GBIG1 ends in a Minkowski universe.

\item $II:~\sigma_{\rm i}<\sigma\leq\sigma_{\rm
c},~\gamma_{\rm i}<\gamma\leq\gamma_{\rm M}$. GBIG1-2 do not exist.
GBIG3 expands to $\mu_{\rm c}$ and then collapses. When
$\sigma=\sigma_{\rm c}$ GBIG3 evolves to a Minkowski universe.

\item $III:~\sigma>\sigma_{\rm c},~\gamma_{\rm i}<\gamma\leq\gamma_{\rm M}$.
GBIG1-2 do not exist. GBIG3 evolves to $h_{\infty}$.

\item $IV:~\sigma_{\rm b}<\sigma\leq\sigma_{\rm i}$ and $~\sigma_{\rm
b}<\sigma\leq\sigma_{\rm c}$; $\gamma\leq\gamma_{\rm M}$ and
$\gamma\leq\gamma_{\rm i}$. GBIG1-2 do not exist. GBIG3 expands to
$\mu_{\rm c}$ and then collapses. When $\gamma=\gamma_{\rm i}$ (and
for $\sigma=\sigma_{\rm i}$) GBIG1-2 both live at ($0,h_{\infty}$).
When $\sigma=\sigma_{\rm c}$ GBIG3 evolves to a Minkowski universe.

\item $V:~\sigma>\sigma_{\rm c},~\gamma\leq\gamma_{\rm i}$.
GBIG1-3 evolve to $h_{\infty}$. When $\gamma=\gamma_{\rm i}$ GBIG1-2
both live at ($0,h_{\infty}$).
\end{itemize}

As we decrease the value of $\phi$, region II in Fig.~\ref{gs20}
shrinks (the point $\sigma_{\rm i}(\gamma_{\rm M})$ becomes
increasingly positive). For $\phi\leq-256/9$ region II no longer
exists. The nature of the solutions in the other regions are
unaffected.

\begin{figure}
\includegraphics[height=3in,width=2.75in,angle=270]{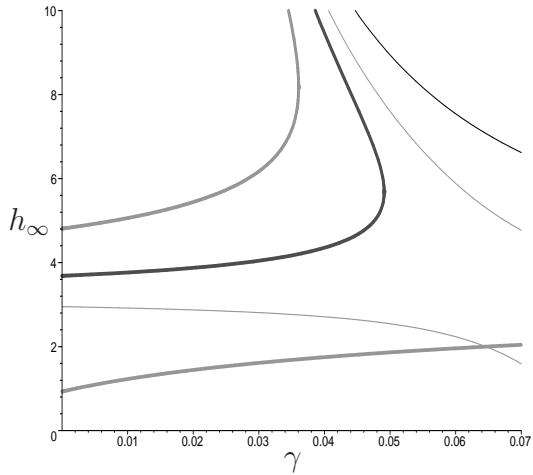}
\rput(-7.4,-3.5){\large $h_{\infty}$}\rput(-3.9,-6.7){\large
$\gamma$}
 \caption{$h_{\infty}$ for solutions in a AdS ($\phi=-20$) bulk.
 The thin-dark line has $\sigma=-10$, thin-light lines have $\sigma=-2$,
 thick-dark lines has $\sigma=2$ and the thick-light lines has $\sigma=10$.}
\label{hinp20}
\end{figure}

There are no bouncing solutions in this case (in fact for any case
with $\phi\leq-9$) as GBIG4 is unallowed and $\gamma_{\rm
M}<\gamma_{\rm m}$ (which rules out the combined solutions). In
Fig.~\ref{hinp20} we show the $h_{\infty}$ results for $\phi=-20$.

\section{Conclusions}

In this work we have looked at the general GBIG model. We have seen
that there is a range of possible dynamics that can be achieved
depending on the parameters in the model. Our main model of interest
is still GBIG1 as this is the one that starts in a finite density
``quiescent'' singularity and evolves to a de Sitter universe. If we
warp the bulk enough, GBIG1 can be allowed to collapse back to its
initial density, provided there is sufficient negative brane
tension. If the bulk is warped enough to allow GBIG1 to collapse
this means that the bouncing cosmology GBIG4 can no longer exist.
The exact nature of the late time dynamics can be changed by
including a non-zero brane tension. A negative brane tension can
reduce the Hubble rate at late time and a positive tension will
increase it. GBIG1 can end in a future ``quiescent'' singularity
with a non-zero density if we have an appropriate (negative) brane
tension present. As the solution of interest is GBIG1 and we want it
to provide the late time acceleration that we are experiencing it
needs to end as a vacuum de Sitter universe. Therefore if there is
some brane tension it must take values $\sigma_{\rm
i}>\sigma>\sigma_{\rm e}$.

GBIG2 also starts in a ``quiescent'' singularity but it
super-accelerates so is un-physical and of little interest.

GBIG3 still starts in an infinite density big-bang but has a number
of possible late time dynamics. In a Minkowski bulk GBIG3 can either
evolve to a Minkowski state, as shown in Ref.~\cite{Brown:2005ug}, a
vacuum de Sitter state or even loiter around $\mu=-\sigma$ before
ending in a vacuum de Sitter state or a ``quiescent'' singularity.
This loitering cosmology is different from that in
Ref.~\cite{Sahni:2004fb}, as we do not require a naked bulk
singularity or a de Sitter bulk. If we warp the bulk GBIG3, will
generally collapse, unless there is sufficient (positive) brane
tension to allow the solution to end in a vacuum de Sitter state.

GBIG4 can only ever exist in a mildly warped bulk with negative
brane tension. So it can be said that GBIG4 is un-physical due to
requirement that $\sigma<0$.

There are a number of bouncing cosmologies within this set-up. There
are the GBIG4 solutions as mentioned above. There are also the
solutions where GBIG4 and 2 match up ($\gamma_{\rm
m}<\gamma\leq\gamma_{\rm M}$ with $\phi_{\rm GBIG4 Lim}<\phi<0$) and
the GBIG2 bouncing cosmologies ($\gamma_{\rm
m}<\gamma\leq\gamma_{\rm M}$ with $-9<\phi\leq\phi_{\rm GBIG4
Lim}$). Each of these have different dynamics so would produce
different evolutionary histories. The solutions that spend time on
the GBIG2 branch will experience phantom like behaviour during this
period.

\[ \]{\bf Acknowledgements:} I am supported by PPARC.
 I thank Roy Maartens and Mariam Bouhmadi-Lopez and for
 very helpful discussions. I would also like to thank Varun Sahni
 for pointing out Ref.~\cite{Shtanov:2002ek}.

\end{document}